\documentclass[final,3p,times, authoryear]{elsarticle}

\usepackage{amsmath,euscript,amssymb,amsfonts,latexsym}
\usepackage[table]{xcolor}
\usepackage{subcaption}
\usepackage{bm}
\usepackage{hyperref}
\usepackage{xspace}
\usepackage{booktabs}
\allowdisplaybreaks

\newcommand{\red}[1]{\textcolor{red}{#1}}

\newcommand{\Rvoid}{\ensuremath{R_{\mathrm{void}}}\xspace}
\newcommand{\delrelk}{\ensuremath{\Delta_\mathrm{rel}\kappa}\xspace}
\newcommand{\gev}{\textsc{Gevolution}\xspace}
\newcommand{\scr}{\textsc{Screening}\xspace}
\newcommand{\gevm}{\texttt{gev}\xspace}
\newcommand{\scrm}{\texttt{scr}\xspace}
\newcommand{\hMpc}{\ensuremath{~h^{-1}\mathrm{Mpc}}\xspace}
\newcommand{\hMsun}{\ensuremath{~h^{-1}\mathrm{M}_\odot}\xspace}

\definecolor{highlight}{RGB}{245,245,245}  

\journal{Physics Letters B}

\begin{document}

\begin{frontmatter}
\title{Tracing Cosmological Signature with Doppler Lensing: Insights from Cosmological Simulations}

\author[first]{Mubtasim Fuad}
\author[second]{Sonia Akter Ema}
\author[first]{Md Rasel Hossen}
\ead{rasel.hossen@juniv.edu}
\affiliation[first]{organization={Department of Physics, Jahangirnagar University},
            addressline={Savar},
            city={Dhaka},
            postcode={1342},
            country={Bangladesh}}
\affiliation[second]{organization={Department of Mathematical \& Physical Sciences, East West University},
            addressline={Aftabnagar},
            city={Dhaka},
            postcode={1212},
            country={Bangladesh}}            

\begin{abstract}
    Doppler lensing, a relativistic effect resulting from the peculiar velocities of galaxies along the line of sight, provides insight into the large-scale structure of the Universe. Relativistic simulations are essential for modeling Doppler lensing because they incorporate gravity and motion in spacetime. We compare two relativistic $N$-body simulation frameworks, \gev and \scr, to calculate Doppler lensing convergence in cosmic voids of different sizes and halos of different masses. Our analysis reveals scale-dependent performance: \scr shows larger differences in small voids (radius range: 15--25\hMpc) with a mean absolute relative difference of 38.5\%, due to linearized dynamics failing in nonlinear regimes. Medium voids (25--35\hMpc) show better agreement (9.5\% mean difference). For large voids (35--45\hMpc), \scr exhibits intermediate differences (16.9\% mean difference) with central instabilities. Moreover, our Doppler convergence analysis with massive halos ($10^{11.5}$--$10^{14}\hMsun$) demonstrates excellent consistency (1.6--3.6\% mean difference). These findings provide clear guidance for simulation choice: \gev is recommended for precision studies critical to $\Lambda$CDM or modified gravity tests, while \scr offers a computationally efficient alternative for relativistic treatments with large catalogs of voids and halos, assisting future astrophysical surveys.
\end{abstract}


\end{frontmatter}




\section{Introduction}
\label{sec:introduction}

The large-scale structure (LSS) of the Universe--a cosmic web of galaxies, filaments, and voids--results from gravitational collapse driven by dark matter and dark energy \citep{peebles1980, springel2005}. Studying LSS evolution is central to cosmology, probing gravity, dark energy, and initial conditions \citep{amendola2018, weltman2020}. In linear regimes with small density contrasts, analytical methods model structure growth \citep{eingorn2022, mukhanov2005, gorbunov2011}. In nonlinear regimes, where density contrasts exceed unity, numerical simulations are essential \citep{davis1985}. Newtonian $N$-body simulations, successful within the $\Lambda$CDM framework \citep{springel2005}, neglect relativistic effects critical at cosmological scales, especially for large peculiar velocities or modified gravity \citep{adamek2016, adamek2016a, eingorn2022}. Next-generation surveys like LSST, Euclid, and SKA will map LSS with high precision, necessitating relativistic frameworks \citep{robertson2017, amendola2018, weltman2020}.

Two relativistic approaches address these needs: \gev and \scr. The \gev code, a relativistic $N$-body tool based on weak-field general relativity, extends Newtonian dynamics but is computationally intensive \citep{adamek2013, adamek2014, adamek2016, adamek2016a}. The \scr approach, using linearized perturbations, is computationally efficient and analytically tractable, suitable for large-scale simulations \citep{eingorn2016, brilenkov2017, eingorn2019, eingorn2022, eingorn2024, eingorn2017, canay2021}. Cosmic voids, underdense regions dominating the Universe's volume, are sensitive probes of dark energy and gravity, with their properties constraining cosmological parameters \citep{2019BAAS...51c..40P, 2022arXiv220107241M, 2012ApJ...761...44S, 2015JCAP...08..028B, 2019MNRAS.490.3573F, 2009ApJ...696L..10L, 2015PhRvD..92h3531P, 2016ApJ...820L...7S}. 

Doppler lensing, caused by galaxy peculiar velocities, probes LSS velocity fields at low redshifts, complementing weak-lensing. It enhances constraints on cosmological parameters and tests $\Lambda$CDM and modified gravity \citep{2013PhRvL.110b1302B, 2014MNRAS.443.1900B, 2017MNRAS.472.3936B, 2020JCAP...12..023H, 2022MNRAS.509.5142H}. The choice of analysing Doppler lensing within cosmic voids is motivated by several key factors. Firstly, being underdense and hence less dynamically complex than halos and clusters, voids serve as cleaner laboratories for isolating the specific relativistic effects encoded in Doppler lensing, which is directly proportional to the line-of-sight peculiar velocity \citep{2013PhRvL.110b1302B}. The velocity flows in voids are more coherent and linear over a larger range of scales compared to the virialized, turbulent motions in clusters \citep{2015PhRvD..92h3531P}. This makes the comparison between the linearized treatment in \scr and the non-linear framework of \gev more interpretable. Secondly, Doppler lensing is particularly pronounced at low redshifts ($z<1$), where the prefactor in Eq.~\eqref{eqn:Dop} is significant, and where upcoming wide-field surveys like Euclid and DESI will provide immense statistical power. While other observables like cluster masses or redshift-space distortions are also valuable, Doppler lensing provides a direct, complementary probe of the velocity field that is exceptionally sensitive to the underlying gravitational framework, especially in the expansive, evolving regions of cosmic voids.
However, to provide a comprehensive comparison between the two frameworks, we extend our analysis beyond cosmic voids to include overdense regions of massive halos. This additional investigation tests the behavior of velocities around both underdense and overdense environments, offering a stronger assessment of the screening approximation's validity across the cosmic web. The halo analysis reveals that \scr performs remarkably well in moderate-density environments, with mean relative differences of only 1.6--3.6\% across five mass bins, demonstrating its reliability for typical large-scale structure studies.

This study compares \gev and \scr in modeling Doppler lensing convergence across a range of void sizes (15--25, 25--35, and 35--45 \hMpc) and halo masses ($10^{11.5}$--$10^{14}\hMsun$) with a focus on their effectiveness \citep{adamek2016} and computational efficiency \citep{eingorn2022,eingorn2024}. Our work reveals multi-scale performance in three distinct void subgroups and also consistent agreement in the halo catalog of various mass ranges. We employ physical and statistical analyses on the methodological performances. Based on these results, we provide guidance for selecting simulation approaches according to scientific requirements: \scr for computationally efficient studies of halos and medium-to-large voids, and \gev for precision analysis in fundamental cosmology tests.

The paper is structured as follows: Section \ref{sec:background} outlines the relevant theoretical background of Doppler lensing, \gev, and \scr frameworks. Section \ref{sec:methodology} details simulation configurations, the method used for halo and void finding, and the Doppler lensing algorithm, and finally, section \ref{sec:results} presents Doppler convergence analyses and results with \gev and \scr.

\section{Background}
\label{sec:background}

\subsection{Doppler Lensing}
\label{sec:doppler-lensing}

Doppler lensing, driven by the peculiar velocities of galaxies, shows changes in their apparent size and magnitude through redshift-space distortions. It offers a complementary probe to weak gravitational lensing, which distorts galaxy shapes due to intervening mass \citep{bonvin2008, 2013PhRvL.110b1302B}. Unlike weak-lensing, Doppler lensing is a local effect of the observed structure, sensitive to the velocity field of source galaxies along the line of sight, making it particularly effective at low redshifts \citep{2014MNRAS.443.1900B, bonvin2017}. The Doppler convergence is expressed as:
\begin{equation}
    \kappa = \left(1 - \frac{1+z_s}{H \chi_s}\right) \frac{\bm{v} \cdot \bm{n}}{c},
    \label{eqn:Dop}
\end{equation}
where $z_s$ is the source redshift, $H$ the Hubble parameter, $\chi_s$ the comoving distance, $\bm{v}$ the peculiar velocity, $\bm{n}$ the unit vector from source to observer, and $c$ the speed of light. If $\bm{v} \cdot \bm{n} > 0$ (motion toward the observer), $\kappa < 0$, causing galaxies to appear smaller and dimmer; the converse holds for receding motion. This sensitivity to velocity fields makes Doppler lensing a powerful tool for studying cosmic voids, where peculiar velocities trace underdense regions, aiding tests of $\Lambda$CDM and modified gravity models \citep{hossen2022}.
In Section \ref{sec:results}, we explore and compare Doppler lensing convergence for two different approaches of $N$-body simulations that are discussed in the subsequent sections. 

\subsection{Gevolution}
\label{sec:gevolution}

The \gev framework, developed by \citet{adamek2016, adamek2016a}, is a relativistic $N$-body simulation tool that extends Newtonian dynamics within a weak-field general relativistic context. It models scalar perturbations by Bardeen potentials ($\Phi$, $\Psi$), vector perturbation by frame-dragging field ($B_i$), and tensor perturbation terms in the perturbed Einstein equations, capturing nonlinear mode couplings critical in high-density regions. The perturbed Friedmann-Lema\^{i}tre-Robertson-Walker (FLRW) metric in the Poisson gauge \citep{mukhanov2005,gorbunov2011} (disregarding tensor modes \citep{eingorn2022}) is:
\begin{align}
ds^2 = a^2 \left[ -(1+2\Psi)d\tau^2 - 2 B_i dx^i d\tau + (1-2\Phi)\delta_{ij} dx^i dx^j \right], \quad i,j=1,2,3, \label{eq:gev-metric}
\end{align}
where $a(\tau)$ is the scale factor of the background depending on conformal time $\tau$, and $x^i$ are comoving coordinates on the spacelike hypersurfaces, the gauge condition $\delta^{ij}B_{i,j}=0$ is satisfied \citep{2012PhRvD..85f3512G,adamek2013,eingorn2022}. \gev's comprehensive treatment of relativistic effects, including large density contrasts, ensures precision in modeling velocity fields for Doppler lensing, but its computational complexity limits efficiency for large-scale statistical analyses.

\subsection{Cosmic Screening}
\label{sec:screening}

The \scr approach, introduced by \citet{eingorn2016, eingorn2022}, streamlines relativistic $N$-body simulations by employing a linearized perturbation framework. By assuming small metric corrections, it derives the first-order gravitational potential through a Helmholtz equation, resulting in a Yukawa-like potential with an exponential cutoff at scales set by the cosmological horizon \citep{brilenkov2017}. This cutoff arises from screening of gravitational interactions due to the background expansion, reducing long-range forces compared to Newtonian gravity. The linearization decouples perturbation orders, omitting nonlinear mode couplings included in \gev, which simplifies computations and achieves approximately 40\% faster performance \citep{eingorn2022}. 
It should be noted that while the same FLRW metric with Poisson gauge is employed, the scalar and vector perturbation functions here handled differently in \scr from those in \gev (Sec.~\ref{sec:gevolution}). In \gev, they include higher-order admixtures, whereas in \scr, they are first-order \citep{eingorn2022}. This separation of first- and second-order perturbations, as advocated by \citet{eingorn2016,brilenkov2017,eingorn2016a}, avoids the complexity of mixing perturbation orders in \gev \citep{adamek2016,eingorn2022}.

This framework's analytical tractability allows for exact solutions to the gravitational potential, facilitating statistical analyses of large-scale structures, such as cosmic voids, in surveys like Euclid and DESI. For Doppler lensing, \scr effectively models velocity fields in medium-to-large voids where linear dynamics dominate, but its neglect of nonlinear terms can lead to inaccuracies in small voids with high density contrasts, where peculiar velocities are more complex. The methodology of the simulations by \scr makes it a valid parallel approach with respect to \gev's conventional method, ideal for balancing computational speed and cosmological accuracy in large-scale studies.

\section{Methodology}
\label{sec:methodology}

\subsection{Methods for N-body simulations}

In this work, we employ the perturbed FLRW metric in Poisson gauge for both relativistic $N$-body approaches \gev \ref{sec:gevolution} and \scr \ref{sec:screening} given by Eq.~\eqref{eq:gev-metric}. We focus on the output of scalar perturbations from our simulations to optimize the computations; also, the vector and tensor perturbations have negligible effects on our context of light propagation in Doppler lensing \citep{lepori2020,ema2022,hossen2022}.

Initial conditions for the \gev and \scr simulations are generated within the standard $\Lambda$CDM framework using the \texttt{CLASS} code \citep{blas2011} at redshift $z_{\text{ini}} = 100$. The primordial power spectrum is defined with a spectral index $n_s = 0.9619$, amplitude $A_s = 2.215 \times 10^{-9}$, and pivot scale $k_{\text{pivot}} = 0.05 \, \text{Mpc}^{-1}$. The simulation uses a grid of $N_{\text{grid}} = 256^3$ in a cubic comoving volume of $(320\hMpc)^3$, yielding a spatial resolution of $1.25\hMpc$. The cosmological parameters are chosen as follows: reduced Hubble constant $h = 0.67556$, physical baryon density $\omega_b = \Omega_b h^2 \approx 0.02204$ (corresponding to $\Omega_b = 0.0483$), physical cold dark matter density $\omega_{\text{CDM}} = \Omega_{\text{CDM}} h^2 \approx 0.12039$ ($\Omega_{\text{c}} = 0.2638$), and effective number of relativistic species $N_{\text{eff}} = N_{\text{ur}} = 3.046$, consistent with Planck 2018 results \citep{aghanim2020}.

Initial conditions are computed using \texttt{CLASS} with a tabulated transfer function in Newtonian gauge at $z_{\text{ini}} = 100$. The particle distribution is based on a homogeneous template (Gadget-2 format) with a tiling factor of 64, and the displacement field is corrected for template pattern convolution. The baryon treatment is set to ``blend'' mode, and the k-domain is ``spherical''. We run 21 realizations for both \gev and \scr simulations, using identical cosmological parameters but different randomly generated seeds. Outputs include particle snapshots (Gadget-2 and $\Phi$ formats) and power spectra at different redshifts. The gravity solver uses General Relativity with an elliptic vector method, a Courant factor of 48.0, and a time step limit of 0.04 in Hubble time units.

\subsection{Methods for Halo and Void Finding}
\label{sec:halo-void-finding}
Dark matter haloes, identified by \textsc{Rockstar} \citep{2013ApJ...762..109B} from particle snapshots of 21 \gev and \scr simulations at $z=0.22$, serve as tracers for both void identification and direct halo analysis. The redshift $z=0.22$ is chosen for Doppler lensing's prominence at low redshifts ($z < 0.3$), where peculiar velocities dominate \citep{2014MNRAS.443.1900B}.

For void analysis, cosmic voids are identified using \textsc{Revolver} \citep{nadathur2019b,nadathur2019}, a 3D void finder based on the \textsc{Zobov} algorithm \citep{2008MNRAS.386.2101N}. \textsc{Zobov} uses Voronoi tessellation to detect density minima from tracer distributions, defining void centers as the largest inscribed empty spheres and computing effective radii as:
\begin{equation}
    R_{\rm void} = \left( \frac{3}{4\pi} V_{\rm void} \right)^{1/3},
    \label{eq:Rv}
\end{equation}
where $V_{\rm void}$ is the total volume of Voronoi cells.

For halo analysis, we select halos in mass bins spanning $\log_{10}(M/\hMsun) \in [11.5, 14.0)$ for direct Doppler convergence computation. The $10^{12}$--$10^{12.5}\hMsun$ range contains the highest halo number density in our simulations. The \textsc{Revolver} algorithm employs a parameter-free approach that provides consistent void detection across our 21 realizations, yielding statistically reliable populations for voids of 15--25, 25--35, and 35--45\hMpc. These size ranges were selected to systematically probe Doppler lensing effects across scales transitioning from nonlinear to linear dynamical regimes.

\subsection{Doppler lensing algorithm}

The Doppler lensing convergence $\kappa$, sensitive to peculiar velocities in cosmic structures, is computed using velocity fields from 21 \gev and \scr simulations, leveraging \gev's relativistic corrections and \scr's linearized dynamics \citep{2014MNRAS.443.1900B}. The Doppler lensing algorithm by \citet{2022MNRAS.509.5142H,hossen2022}, optimized for both underdense and overdense regions, integrates velocities along the line of sight, producing radial convergence profiles for voids and halos. Following Eq.~\eqref{eqn:Dop}, the computation involves the following steps:

\begin{itemize}
    \item \textbf{Generating particle snapshots:} We generate particle snapshots from 21 \gev and 21 \scr simulations, each with unique initial conditions but identical $\Lambda$CDM cosmological parameters.
    
    \item \textbf{Finding cosmic structures:} Using \textsc{Rockstar}, we detect dark matter haloes, extracting positions, masses, and peculiar velocities as tracers. For void analysis, \textsc{Revolver} identifies cosmic voids, providing their positions, radii, and volumes, enabling void-centric Doppler lensing analysis.
    
    \item \textbf{Targeting cosmic structures:} 
    \begin{itemize}
        \item \textbf{Voids:} To mitigate sampling bias, we subdivide voids into three radius ranges: $R_{\rm void} \in [15, 25)\hMpc$, $[25, 35)\hMpc$, and $[35, 45)\hMpc$. These bins probe dynamic velocity fields, from nonlinear to linear regimes.
        \item \textbf{Halos:} We analyze halos in different mass bins: $\log_{10}(M/\hMsun) \in [11.5, 14.0)$, providing a complementary analysis of overdense regions.
    \end{itemize}
    
    \item \textbf{Computing impact parameter:} 
    \begin{itemize}
        \item \textbf{For voids:} We calculate the impact parameter $b_{\rm void} = (R/R_{\rm void})\cos \Theta$, where $R$ is the distance from the void center to halo centers, and $\Theta$ is the angle between the observer-to-void-center direction and the void-center-to-halo direction \citep{2022MNRAS.509.5142H}.
        \item \textbf{For halos:} We compute $b_{\rm halo} = (R/R_{\rm halo})\cos \Theta$, where $R_{\rm halo}$ is the halo virial radius, and $\Theta$ is defined similarly with respect to the halo center.
    \end{itemize}
    
    \item \textbf{Computing and stacking the Doppler convergence:} We compute the dot product of normalized halo velocities ($\bm{v}/c$) and the direction vector ($\bm{n}$) from observer to halo, capturing velocity gradients. The Doppler convergence $\kappa$ is obtained by multiplying this with the redshift-dependent prefactor in Eq.~\eqref{eqn:Dop}. Finally, we stack $\kappa$ values across 21 realizations for each void radius bin and halo mass bin, producing robust profiles to evaluate \gev's precision and \scr's efficiency across different cosmic environments.
\end{itemize}

This algorithm, enhanced by multi-realization stacking, ensures accurate velocity field mapping in both voids and halos, enabling precise comparisons of relativistic effects in Doppler lensing for cosmological surveys like Euclid and DESI.

\subsection{Comparison Methodology}
We quantify differences in Doppler convergence profiles between \gev and \scr using different metrics: raw convergence difference $\Delta\kappa$, and relative difference in symmetrized form $\delrelk$:
\begin{gather}
    \Delta\kappa = \kappa_{\scrm} - \kappa_{\gevm}
    \label{eqn:delta-kappa} \\
    \delrelk = \frac{\Delta\kappa}{\langle |\kappa| \rangle} =  \frac{\Delta\kappa}{\frac{1}{2}(|\kappa_{\scrm}| + |\kappa_{\gevm}|)}
    \label{eqn:delta-rel-kappa}
\end{gather}
Also, the statistical significance in each bin is assessed using $z$-scores:
\begin{equation}
    z = \frac{\Delta\kappa}{\sqrt{\frac{\sigma^2_{\scrm}}{N_{\scrm}} + \frac{\sigma^2_{\gevm}}{N_{\gevm}}}}
    \label{eqn:z-score}
\end{equation}
Here, $N_{\gevm}$ and $N_{\scrm}$ are galaxy counts per bin of normalized impact parameter for \gev and \scr, respectively. Thresholds for $z$-score are adjusted for comparisons along the chosen three ranges of void sizes using Bonferroni correction (e.g., $|z|>2.91$, $p<3.57\times10^{-3}$ for 15--25\hMpc). 
Binning parameters are optimized using total void counts across 21 realizations to balance resolution and statistical reliability for void radial ranges 15--25, 25--35, and 35--45\hMpc (Table~\ref{tab:void_optimization}).
The same metrics (Eqs.~\eqref{eqn:delta-kappa}--\eqref{eqn:z-score}) are applied to both void and halo analyses, enabling consistent comparison across cosmic environments.

\section{Results and Analysis}
\label{sec:results}

This section analyzes Doppler lensing convergence profiles from \gev and \scr across voids of 15--25, 25--35, and 35--45\hMpc and halos spanning $10^{11.5}$--$10^{14}\hMsun$, quantifying performance differences using 21 realizations. The void results show a pattern over different sizes of void radius: largest differences for small voids (15–25\hMpc), smaller differences for medium voids (25–35\hMpc), and intermediate differences for large voids (35–45\hMpc). This is complemented by the strong consistency found in halo environments. Together, these findings provide methodological guidance for choosing simulation approaches based on the specific cosmic structures being studied, as detailed below.


\begin{figure*}[!htbp]
    \centering
    \begin{subfigure}{0.3\textwidth}
        \centering
        \includegraphics[width=\linewidth]{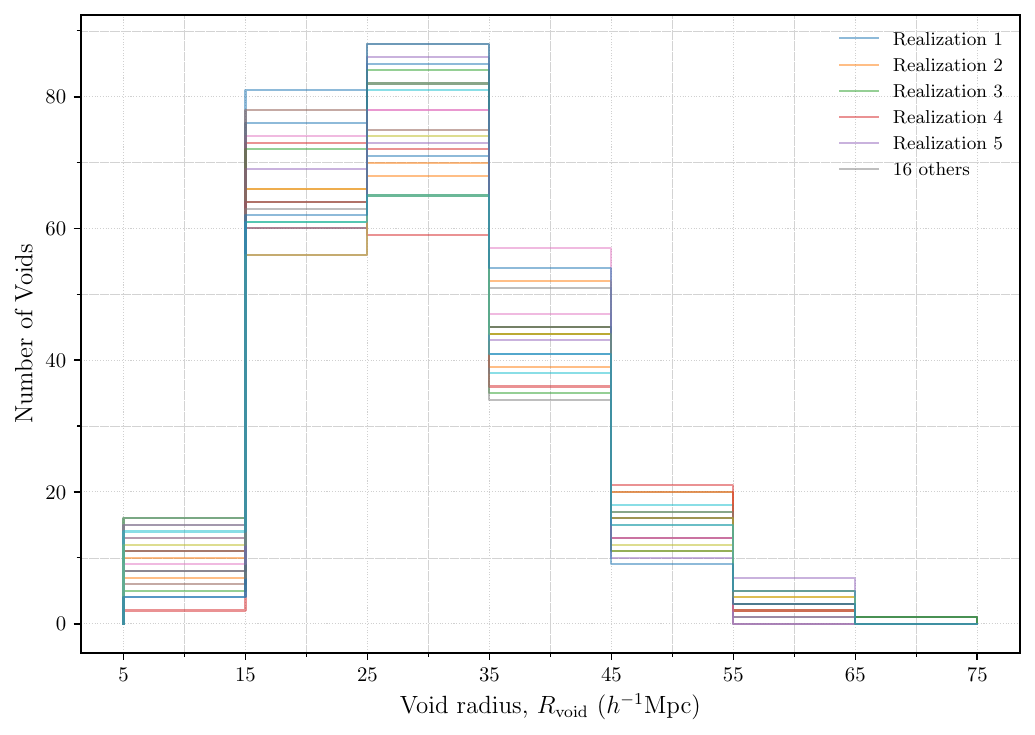}
        \caption{\gev void population.}
        \label{fig:void_count_gev}
    \end{subfigure}
    \begin{subfigure}{0.3\textwidth}
        \centering
        \includegraphics[width=\linewidth]{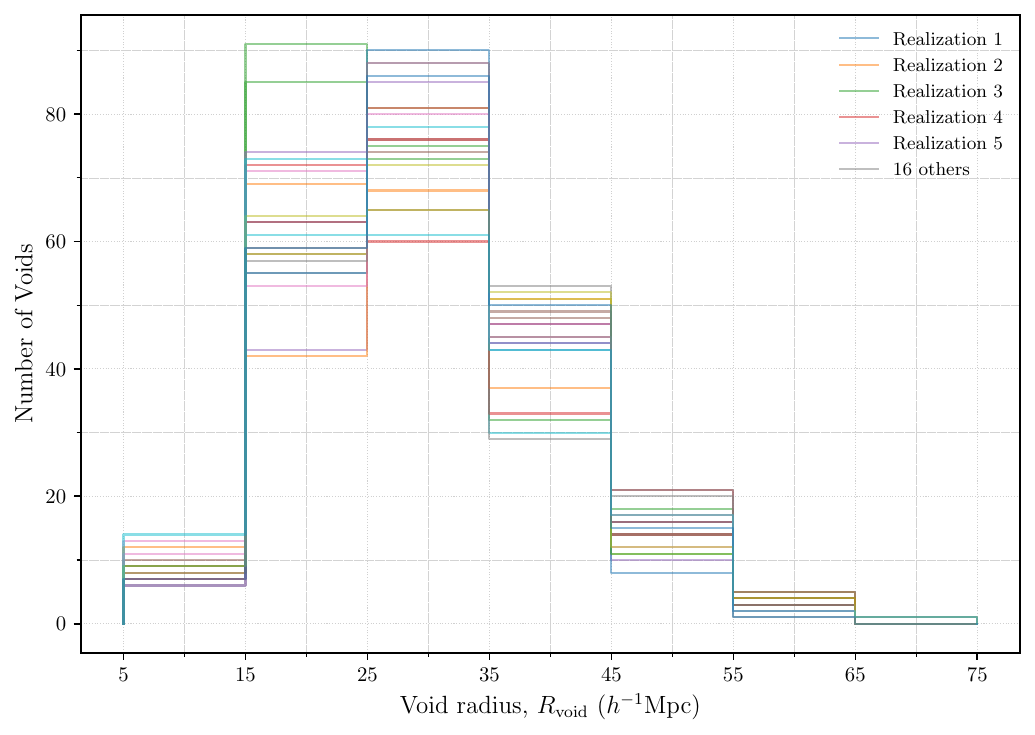}
        \caption{\scr void population.}
        \label{fig:void_count_scr}
    \end{subfigure}
    \begin{subfigure}{0.3\textwidth}
        \centering
        \includegraphics[width=\linewidth]{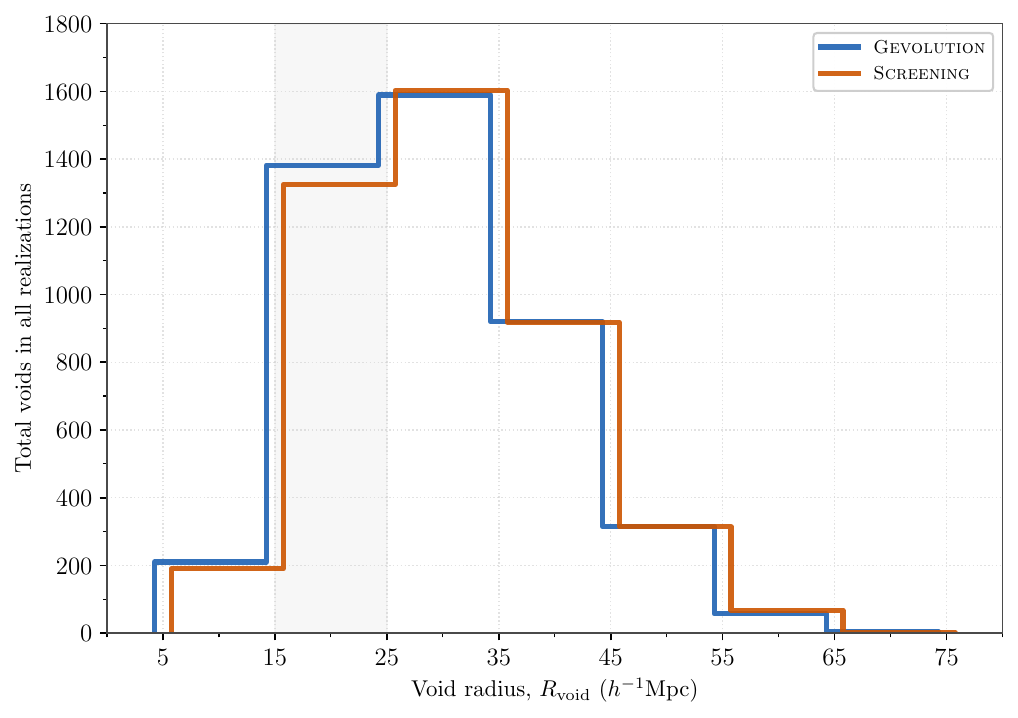}
        \caption{Total void comparison.}
        \label{fig:void_count_total}
    \end{subfigure}
    \caption{Histogram of void radius distributions from 21 realizations of \gev (blue) and \scr (orange) simulations, identified using \textsc{Revolver}. The 15--25\hMpc range shows a 4.2\% deficit in \scr voids compared to \gev, while larger voids (25--35 and 35--45\hMpc) are nearly equivalent (see Table~\ref{tab:void_optimization}).}
    \label{fig:void-radi-count}
\end{figure*}

Void populations, summarized in Table \ref{tab:void_optimization}, underpin convergence analysis. \scr undercounts small voids (15--25\hMpc) by 4.2\% (1325 vs. 1382 for \gev), reflecting linearized dynamics' limitations in nonlinear regimes (reliability score 0.92). Medium voids (25--35\hMpc) show a 0.8\% surplus (1603 vs. 1589), and large voids (35--45\hMpc) are balanced (918 vs. 920, 0.2\% deficit), with reliability scores of 0.95 and 0.88, respectively. Figure \ref{fig:void-radi-count} visualizes these distributions, indicating \scr's suitability for larger voids and \gev's precision for smaller ones.

\begin{table*}[!htbp]
    \centering
    \begin{tabular}{ccccccc}
    \toprule
    \textbf{Range} & \textbf{\gev} & \textbf{\scr} & \textbf{Relative} & \textbf{Optimized} & \textbf{Optimized} & \textbf{Reliability} \\
    \textbf{(\hMpc)} & \textbf{Total Voids} & \textbf{Total Voids} & \textbf{Diff (\%)} & \textbf{Bins} & \textbf{Min Count} & \textbf{Score} \\
    \midrule
    $[5, 15)$  & 210 & 190 & $-10.0$ & 5 & 7 & 0.21 \\
    \rowcolor{highlight}
    $[15, 25)$ & 1382 & 1325 & $-4.2$ & 14 & 18 & 0.92 \\
    \rowcolor{highlight}
    $[25, 35)$ & 1589 & 1603 & $+0.8$ & 16 & 19 & 0.95 \\
    \rowcolor{highlight}
    $[35, 45)$ & 920 & 918 & $-0.2$ & 12 & 15 & 0.88 \\
    $[45, 55)$ & 315 & 314 & $-0.3$ & 4 & 15 & 0.50 \\
    $[55, 65)$ & 57 & 68 & $+17.6$ & 2 & 5 & 0.14 \\
    $[65, 75)$ & 6 & 3 & $-66.7$ & 1 & 1 & 0.12 \\
    \bottomrule
    \end{tabular}
    \caption{Void population and optimization parameters from 21 \gev and \scr realizations. Highlighted ranges (reliability score $>0.7$) are selected for Doppler convergence analysis.}
    \label{tab:void_optimization}
\end{table*}

\begin{figure*}[!htbp]
    \centering
    \begin{subfigure}{0.4\textwidth}
        \centering
        \includegraphics[width=\linewidth]{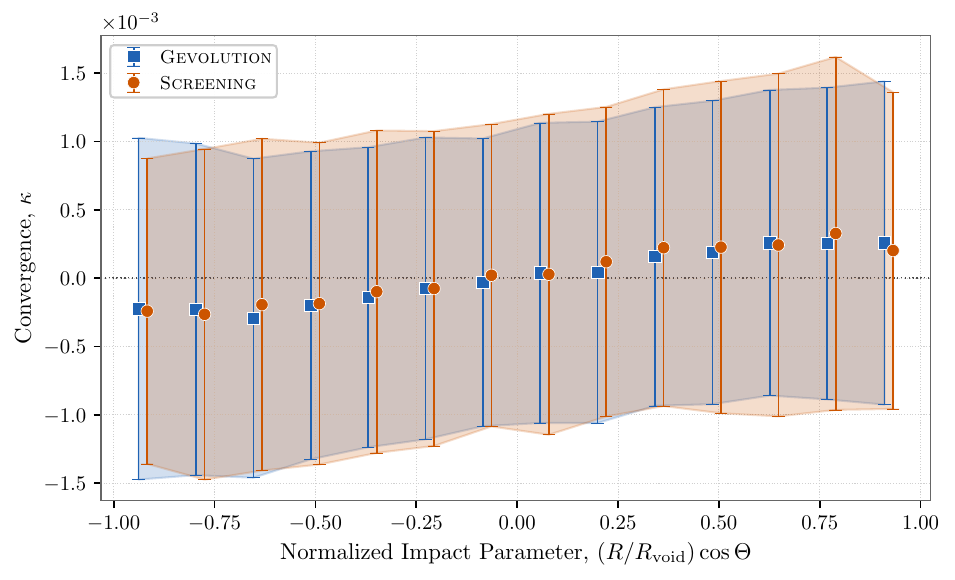}
        \caption{$\kappa$, 15--25\hMpc voids.}
        \label{fig:k_shaded_15-25}
    \end{subfigure}
    \begin{subfigure}{0.4\textwidth}
        \centering
        \includegraphics[width=\linewidth]{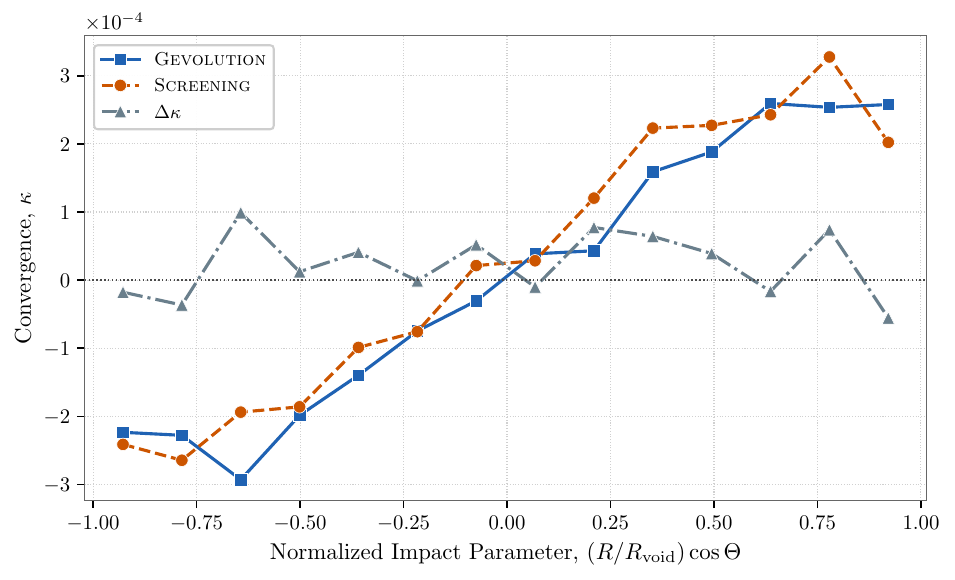}
        \caption{$\kappa$ and $\Delta\kappa$, 15--25\hMpc voids.}
        \label{fig:k_delta_k_15-25}
    \end{subfigure}
    \begin{subfigure}{0.4\textwidth}
        \centering
        \includegraphics[width=\linewidth]{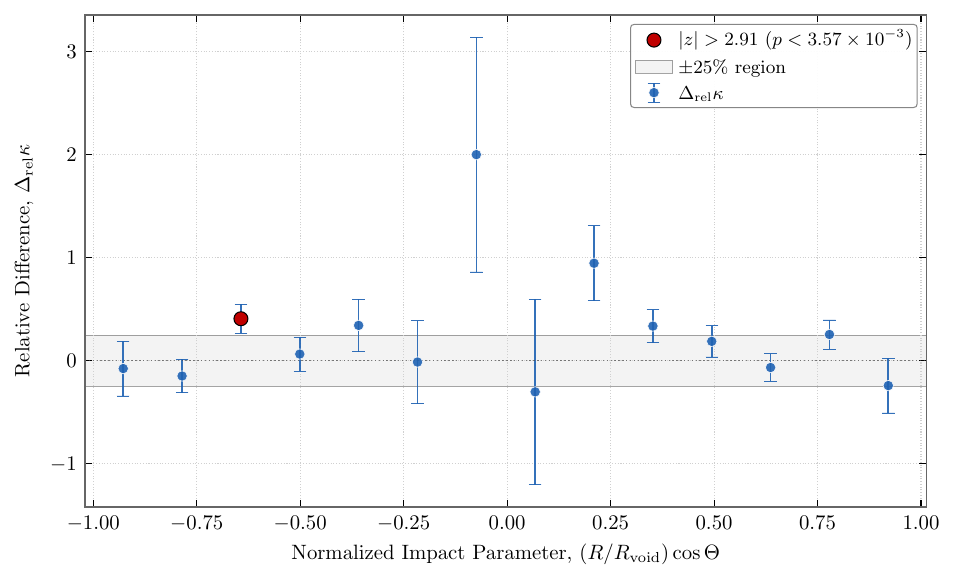}
        \caption{$\delrelk$, 15--25\hMpc voids.}
        \label{fig:delta_rel_k_15-25}
    \end{subfigure}
    \begin{subfigure}{0.4\textwidth}
        \centering
        \includegraphics[width=\linewidth]{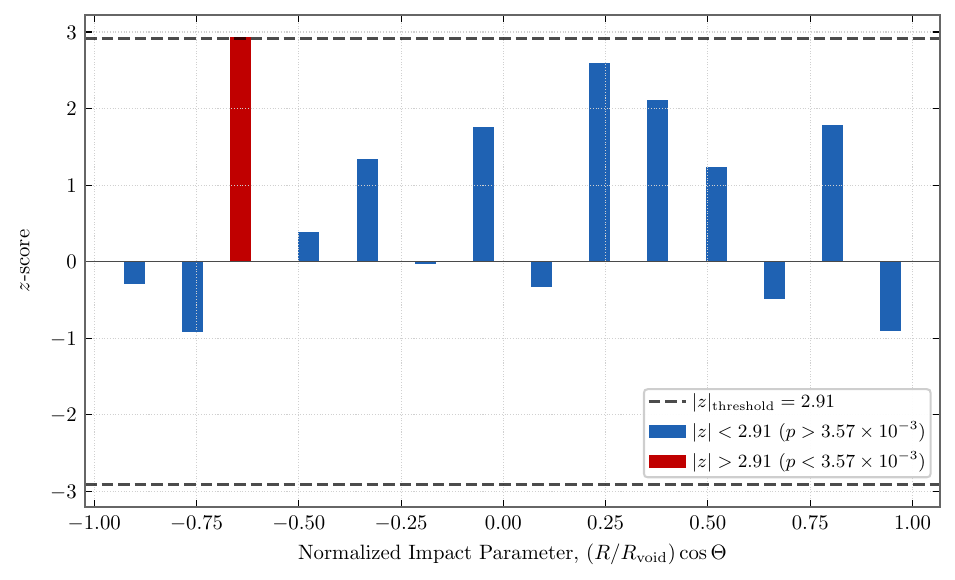}
        \caption{$z$-score, 15--25\hMpc voids.}
        \label{fig:z_score_15-25}
    \end{subfigure}
    \caption{Doppler lensing convergence for 15--25\hMpc voids from 21 \gev and \scr realizations, showing \scr's biases in nonlinear regimes.}
    \label{fig:k_15-25}
\end{figure*}
\begin{figure*}[!htbp]
    \centering
    \begin{subfigure}{0.4\textwidth}
        \centering
        \includegraphics[width=\linewidth]{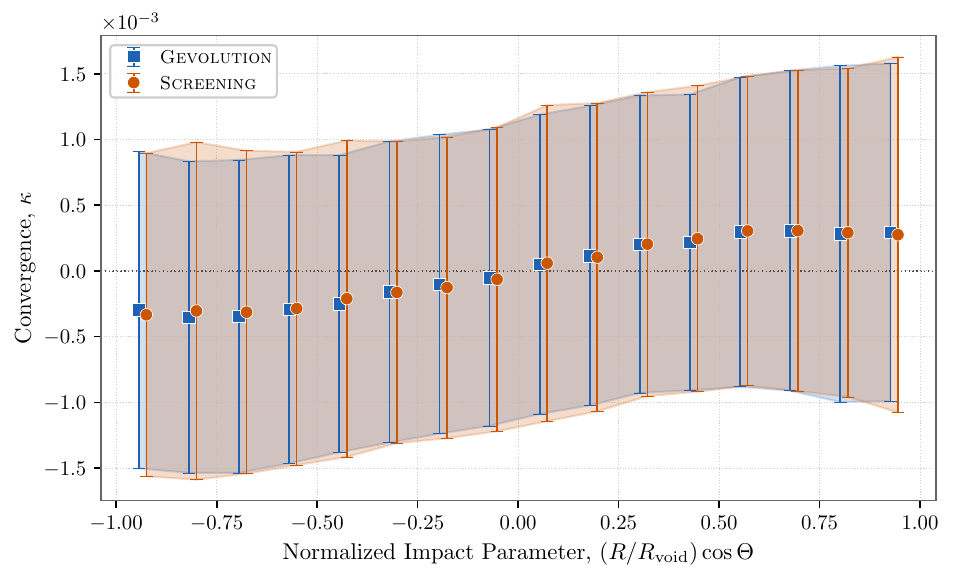}
        \caption{$\kappa$, 25--35\hMpc voids.}
        \label{fig:k_shaded_25-35}
    \end{subfigure}
    \begin{subfigure}{0.4\textwidth}
        \centering
        \includegraphics[width=\linewidth]{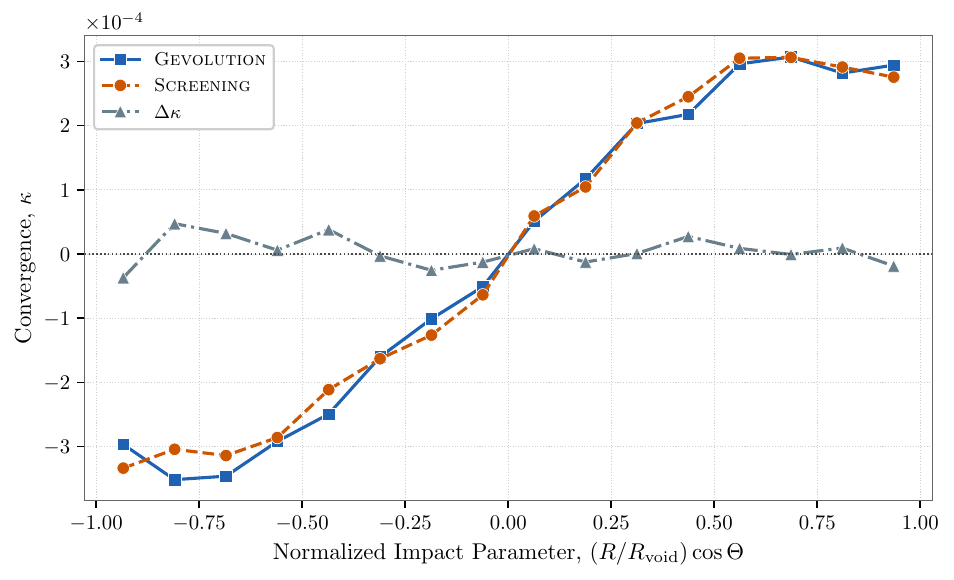}
        \caption{$\kappa$ and $\Delta\kappa$, 25--35\hMpc voids.}
        \label{fig:k_delta_k_25-35}
    \end{subfigure}
    \begin{subfigure}{0.4\textwidth}
        \centering
        \includegraphics[width=\linewidth]{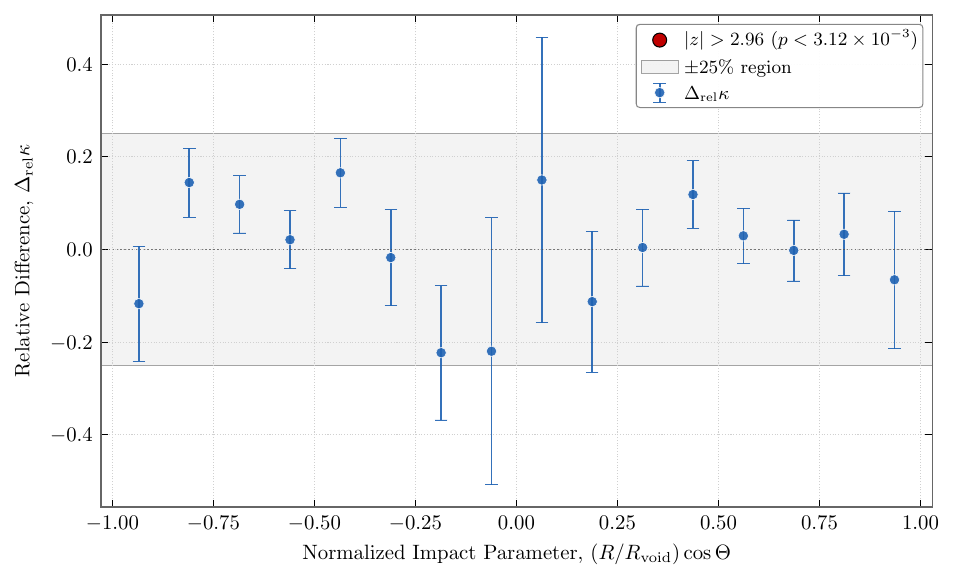}
        \caption{$\delrelk$, 25--35\hMpc voids.}
        \label{fig:delta_rel_k_25-35}
    \end{subfigure}
    \begin{subfigure}{0.4\textwidth}
        \centering
        \includegraphics[width=\linewidth]{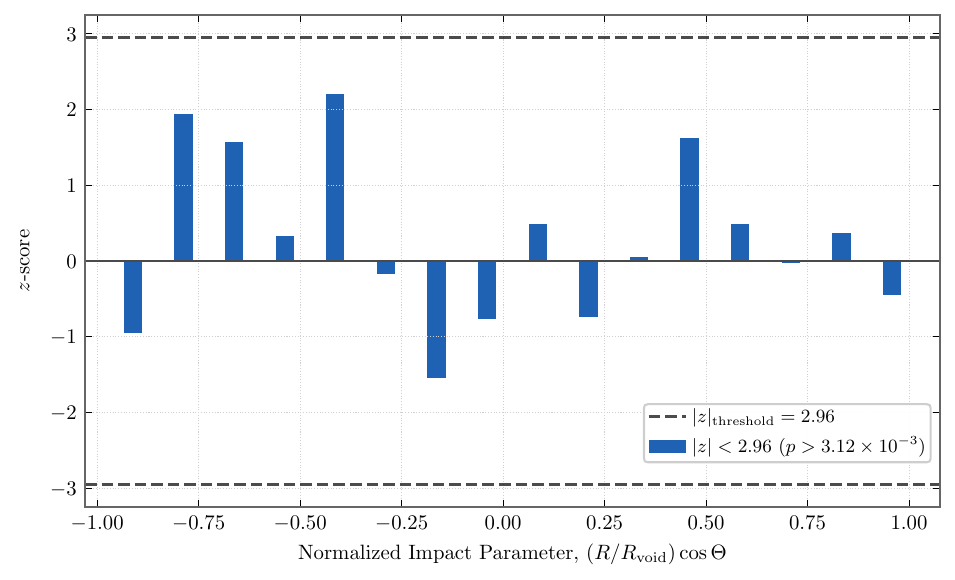}
        \caption{$z$-score, 25--35\hMpc voids.}
        \label{fig:z_score_25-35}
    \end{subfigure}
    \caption{Doppler lensing convergence for 25--35\hMpc voids, showing \scr's reliable performance for medium voids.}
    \label{fig:k_25-35}
\end{figure*}
\begin{figure*}[!htbp]
    \centering
    \begin{subfigure}{0.4\textwidth}
        \centering
        \includegraphics[width=\linewidth]{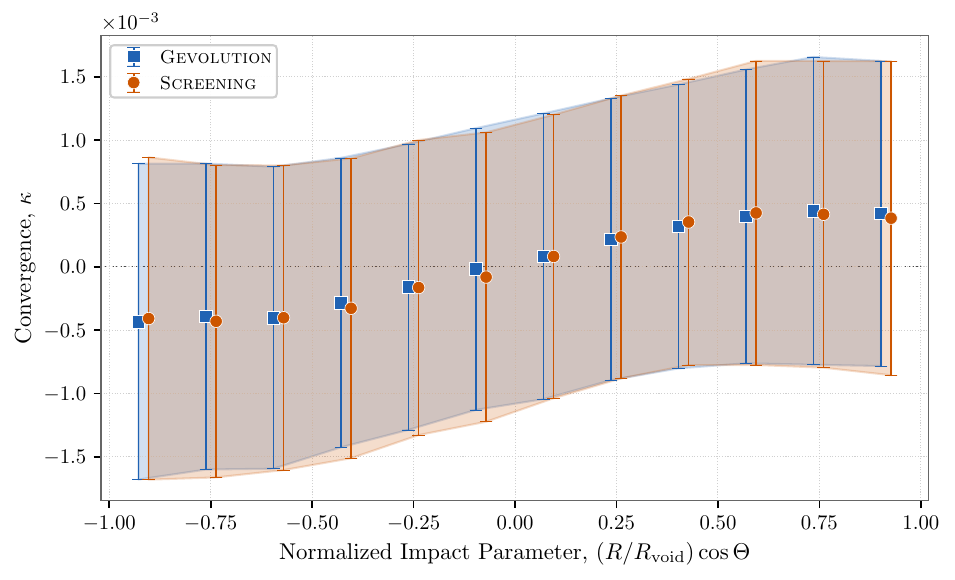}
        \caption{$\kappa$, 35--45\hMpc voids.}
        \label{fig:k_shaded_35-45}
    \end{subfigure}
    \begin{subfigure}{0.4\textwidth}
        \centering
        \includegraphics[width=\linewidth]{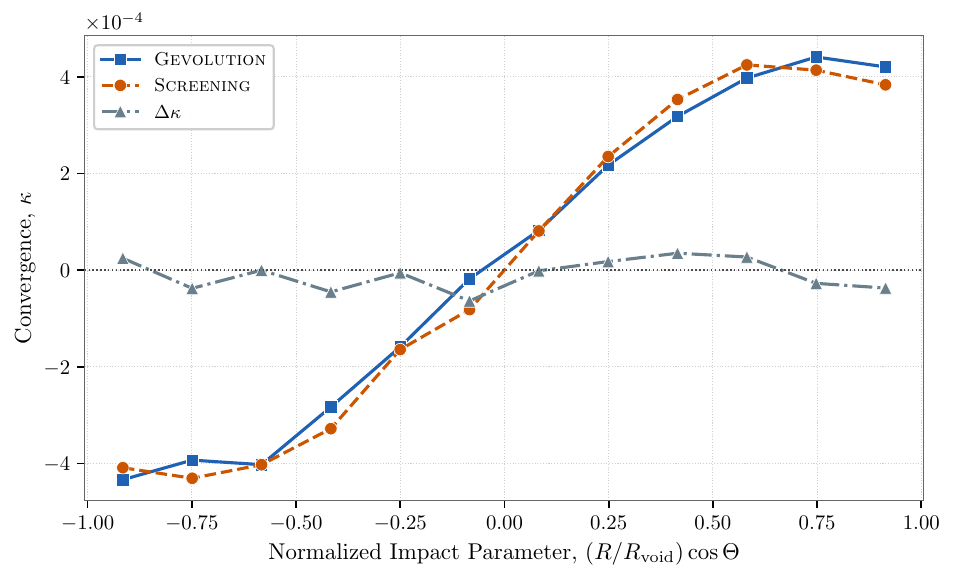}
        \caption{$\kappa$ and $\Delta\kappa$, 35--45\hMpc voids.}
        \label{fig:k_delta_k_35-45}
    \end{subfigure}
    \begin{subfigure}{0.4\textwidth}
        \centering
        \includegraphics[width=\linewidth]{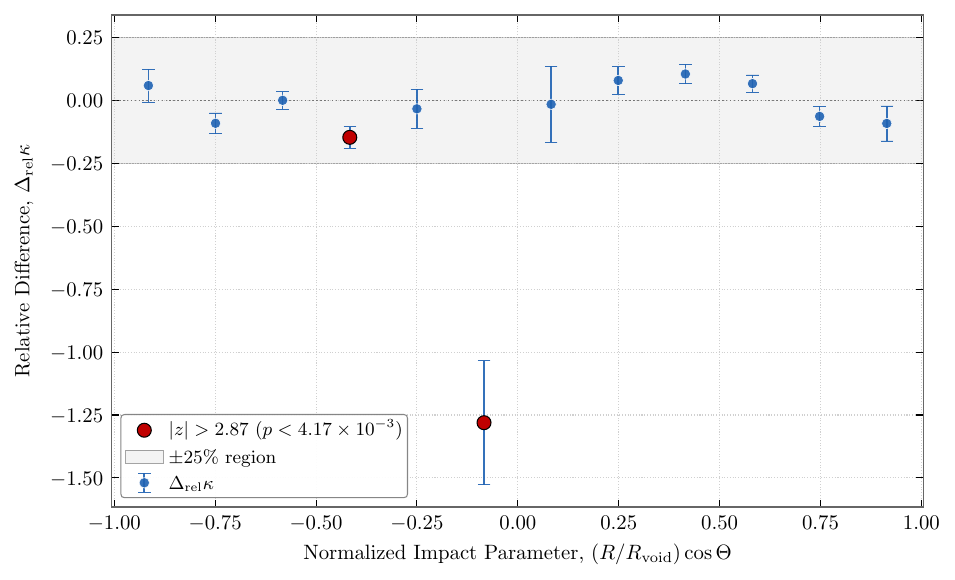}
        \caption{$\delrelk$, 35--45\hMpc voids.}
        \label{fig:delta_rel_k_35-35}
    \end{subfigure}
    \begin{subfigure}{0.4\textwidth}
        \centering
        \includegraphics[width=\linewidth]{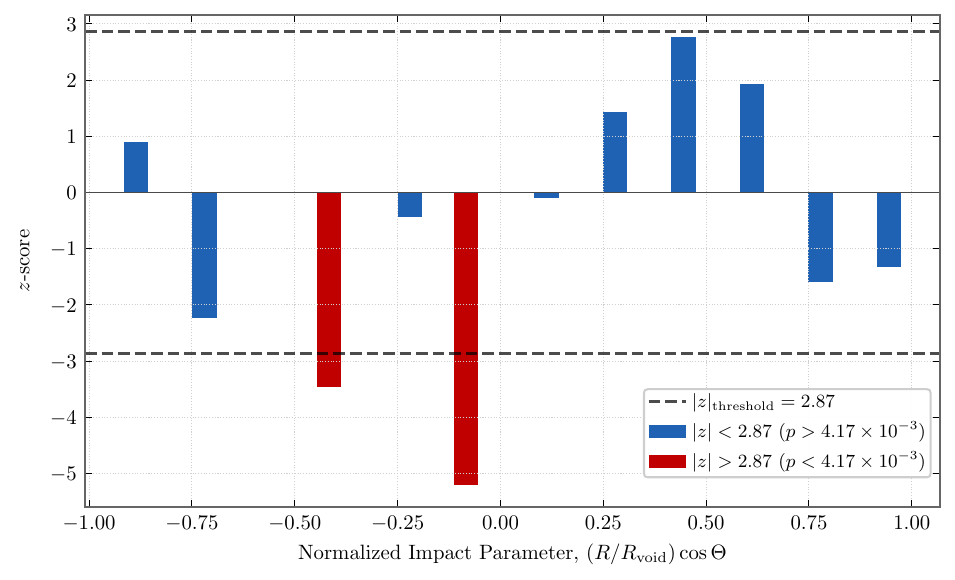}
        \caption{$z$-score, 35--45\hMpc voids.}
        \label{fig:z_score_35-45}
    \end{subfigure}
    \caption{Doppler lensing convergence for 35--45\hMpc voids, indicating \scr's moderate errors due to central instabilities.}
    \label{fig:k_35-45}
\end{figure*}

Convergence $\kappa$, binned by normalized impact parameter $b = (R/\Rvoid)\cos\Theta$, reveals \scr's performance relative to \gev's (Figs.~\ref{fig:k_15-25}--\ref{fig:k_35-45}). Below, we analyze profiles for each void size, focusing on significant discrepancies.

For 15--25\hMpc voids, \scr underestimates approaching flows ($\kappa < 0$) by 9.6\% and overestimates receding flows ($\kappa > 0$) by 15.9\%, yielding a mean $|\delrelk| = 38.5\%$ across 14 bins (50.0\% within $\pm 25\%$). Figure \ref{fig:k_shaded_15-25} shows systematic deviations, with peak $\Delta\kappa = 5.18 \times 10^{-5}$ at $b = -0.07$. A significant bin at $b = -0.64$ ($\delrelk = 40.7\%$, $z = 2.93$, $p = 0.00339$) underscores \scr's nonlinear regime limitations (Fig.~\ref{fig:delta_rel_k_15-25}).

For 25--35\hMpc voids, \scr shows closer alignment with \gev, with a mean $|\delrelk| = 9.5\%$ and minimal $\pm 1.9\%$ bias across 16 bins (100\% within $\pm 25\%$). Figure \ref{fig:k_shaded_25-35} demonstrates this closer agreement, with small $\Delta\kappa = 2.53 \times 10^{-5}$ at $b = -0.19$ (Fig.~\ref{fig:k_delta_k_25-35}). No bins exceed $|z| > 2.96$, indicating \scr's better performance in moderate-density regimes where linear approximations are more applicable (Fig.~\ref{fig:delta_rel_k_25-35}). Specifically, the $9.5\%$ difference in medium voids indicates our expectation that this systematic error is sub-dominant relative to the statistical uncertainties projected for upcoming Doppler lensing analyses (e.g., Euclid/DESI). A $9.5\%$ difference in the stacked signal on these large scales is anticipated to result in a cosmological inference that agrees at the percent level, as this aligns with the required error budget for $\sim$1\% precision on parameters like the cosmic growth rate $\sigma_8$ when measured using velocity-field probes \citep{2022MNRAS.510.2980A}. This essential distinction validates the utility of screening for efficient statistical analysis across medium scales, whereas the $38.5\%$ difference clearly violates this requirement, demanding the precision of \gev on smaller, non-linear scales.

For 35--45\hMpc voids, \scr shows moderate errors, underestimating approaching flows by 24.9\% and overestimating receding flows by 1.3\%, with a mean $|\delrelk| = 16.9\%$ across 12 bins (91.7\% within $\pm 25\%$). Figure \ref{fig:k_shaded_35-45} highlights central deviations, with $\Delta\kappa = 6.37 \times 10^{-5}$ at $b = -0.08$ (Fig.~\ref{fig:k_delta_k_35-45}). Significant bins at $b = -0.42$ ($\delrelk = -14.7\%$, $z = -3.47$) and $b = -0.08$ ($\delrelk = -128.1\%$, $z = -5.20$) indicate boundary and central instabilities (Fig.~\ref{fig:delta_rel_k_35-35}).

The triphasic pattern (Table~\ref{tab:summary}) reflects voids' size-dependent effects. Small voids (15--25\hMpc) show substantially larger differences from \gev ($|\delrelk| = 38.5\%$, with 7.1\% of bins statistically significant) due to \scr's neglect of nonlinear mode couplings in high-density-contrast regimes, necessitating \gev's precision in these environments. Medium voids (25--35\hMpc) exhibit closer agreement ($|\delrelk| = 9.5\%$, 0\% significant bins), as \scr's linear approximations perform better in moderate-density environments. Large voids (35--45\hMpc) display intermediate differences ($|\delrelk| = 16.9\%$, 16.7\% significant bins) from \scr's limitations in capturing velocity gradients and central instabilities (e.g., $\delrelk = -128.1\%$ at $b = -0.08$). \scr's $\sim$40\% faster computation supports its use for medium-to-large void studies, while \gev provides essential accuracy for small voids. These dynamic differences inform the choice of simulation method based on the specific void properties and scientific objectives.

\begin{table*}[!htbp]
    \centering
    \begin{tabular}{cccccccc}
        \toprule
        \textbf{Void Radius Range} & \textbf{Relative Diff.} & \textbf{Reliability} & \textbf{Mean} & \textbf{Within $|\delrelk|\le0.25$} & \textbf{Mean} \\
        (\!\hMpc) & \textbf{in Voids (\%)} & \textbf{Score} & \textbf{$|\delrelk|$} & \textbf{(\%)} & \textbf{$|z|$} \\
        \midrule
        $[15, 25)$ & $-4.2$ & 0.92 & 0.385 & 50.0 & 1.22 \\
        $[25, 35)$ & $+0.8$ & 0.95 & 0.095 & 100.0 & 0.86 \\
        $[35, 45)$ & $-0.2$ & 0.88 & 0.169 & 91.7 & 1.78 \\
        \bottomrule
    \end{tabular}        
    \caption{Summary of void population and convergence analysis from 21 realizations, showing small voids' large discrepancies (38.5\%), medium voids' agreement (9.5\%), and large voids' moderate errors (16.9\%).}
    \label{tab:summary}
\end{table*}

\begin{figure*}[!htbp]
    \centering
    \begin{subfigure}{0.4\textwidth}
        \centering
        \includegraphics[width=\linewidth]{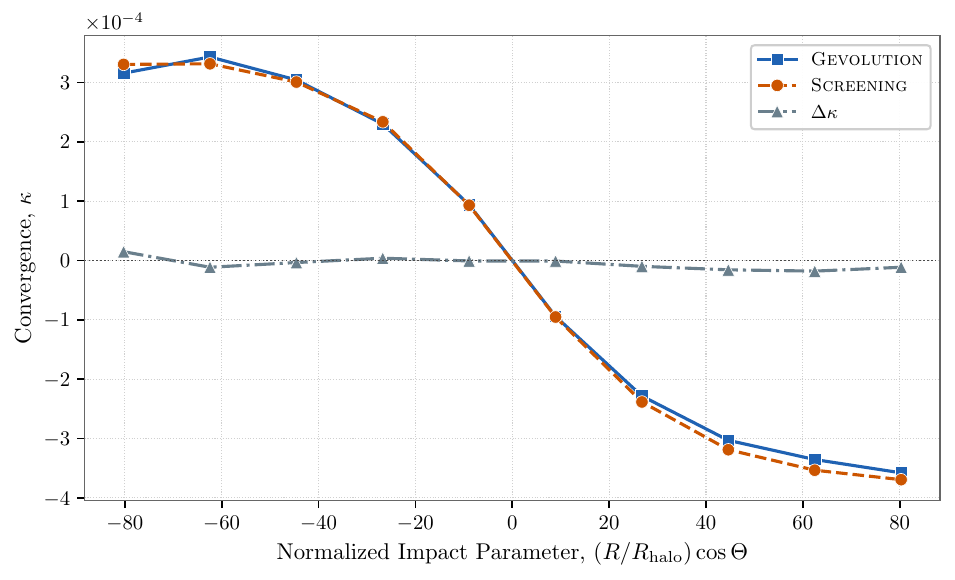}
        \caption{$\kappa$ and $\Delta\kappa$, $10^{11.5}$--$10^{12.0}\hMsun$ halos.}
        \label{fig:k_delta_k_halo_11.5-12.0}
    \end{subfigure}
    \begin{subfigure}{0.4\textwidth}
        \centering
        \includegraphics[width=\linewidth]{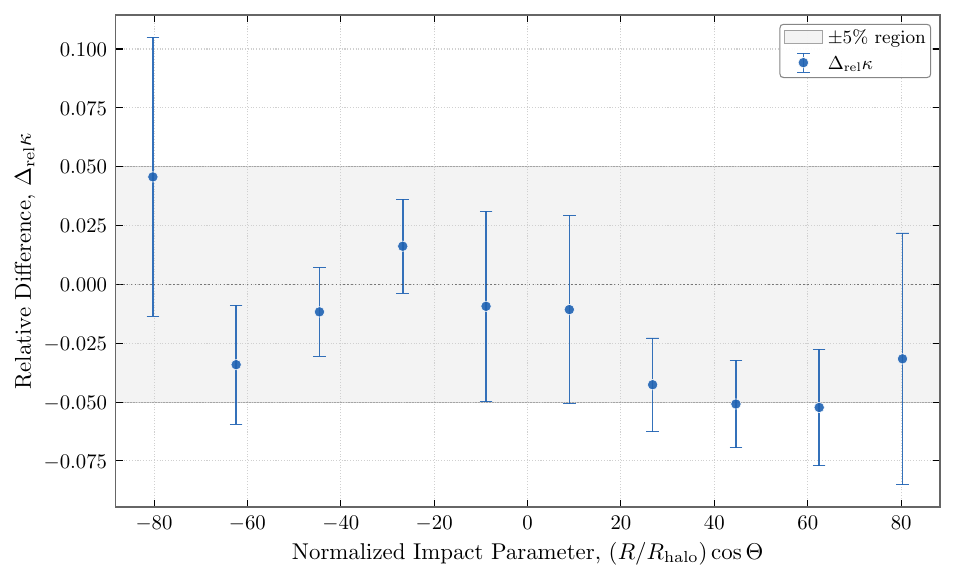}
        \caption{$\delrelk$, $10^{11.5}$--$10^{12.0}\hMsun$ halos.}
        \label{fig:delta_rel_k_halo_11.5-12.0}
    \end{subfigure}
    
    
    \begin{subfigure}{0.4\textwidth}
        \centering
        \includegraphics[width=\linewidth]{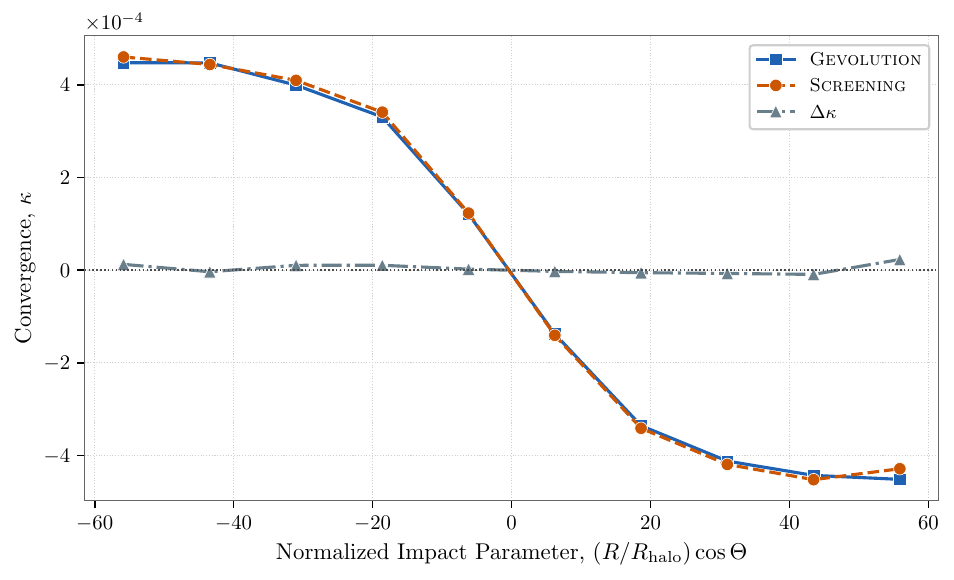}
        \caption{$\kappa$ and $\Delta\kappa$, $10^{12.0}$--$10^{12.5}\hMsun$ halos.}
        \label{fig:k_delta_k_halo_12.0-12.5}
    \end{subfigure}
    \begin{subfigure}{0.4\textwidth}
        \centering
        \includegraphics[width=\linewidth]{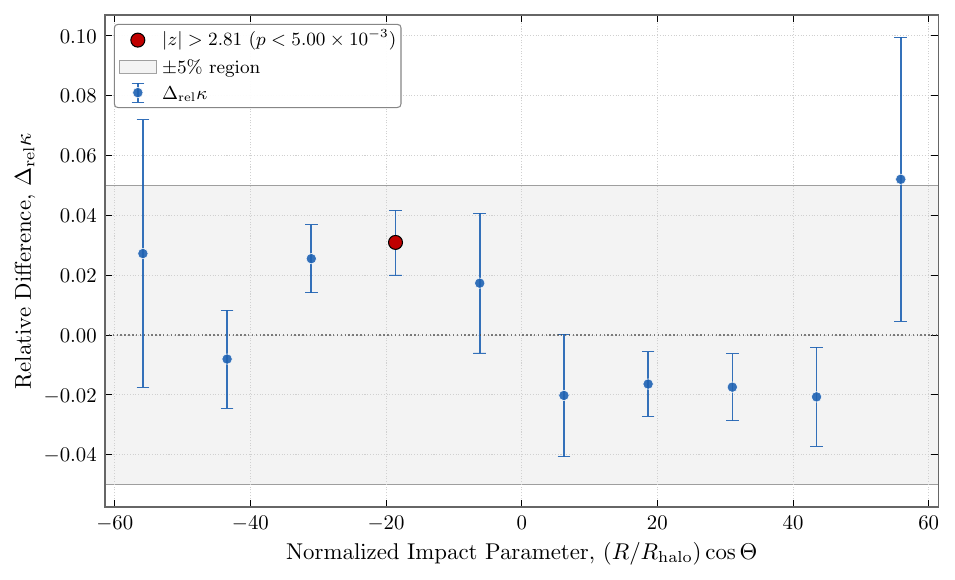}
        \caption{$\delrelk$, $10^{12.0}$--$10^{12.5}\hMsun$ halos.}
        \label{fig:delta_rel_k_halo_12.0-12.5}
    \end{subfigure}
    
    
    \begin{subfigure}{0.4\textwidth}
        \centering
        \includegraphics[width=\linewidth]{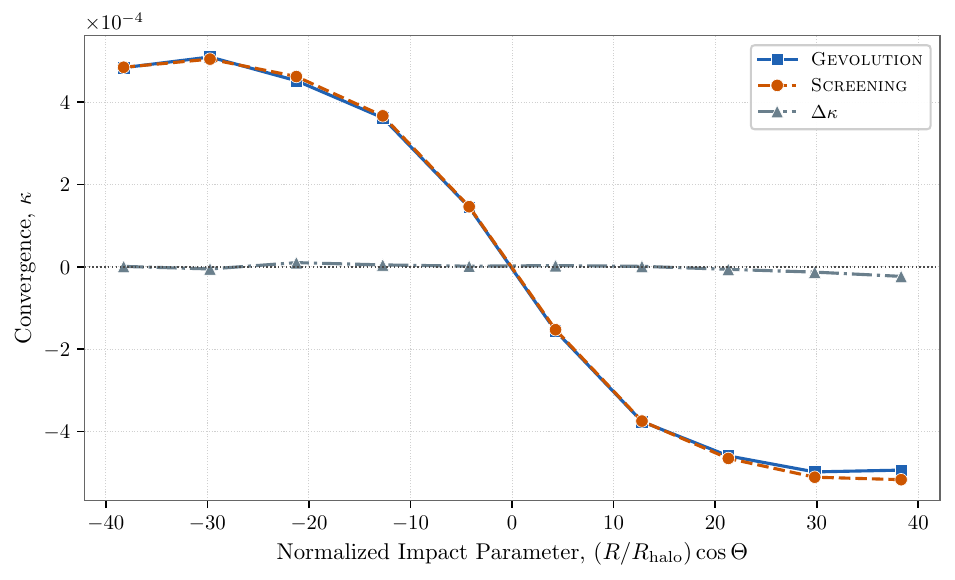}
        \caption{$\kappa$ and $\Delta\kappa$, $10^{12.5}$--$10^{13.0}\hMsun$ halos.}
        \label{fig:k_delta_k_halo_12.5-13.0}
    \end{subfigure}
    \begin{subfigure}{0.4\textwidth}
        \centering
        \includegraphics[width=\linewidth]{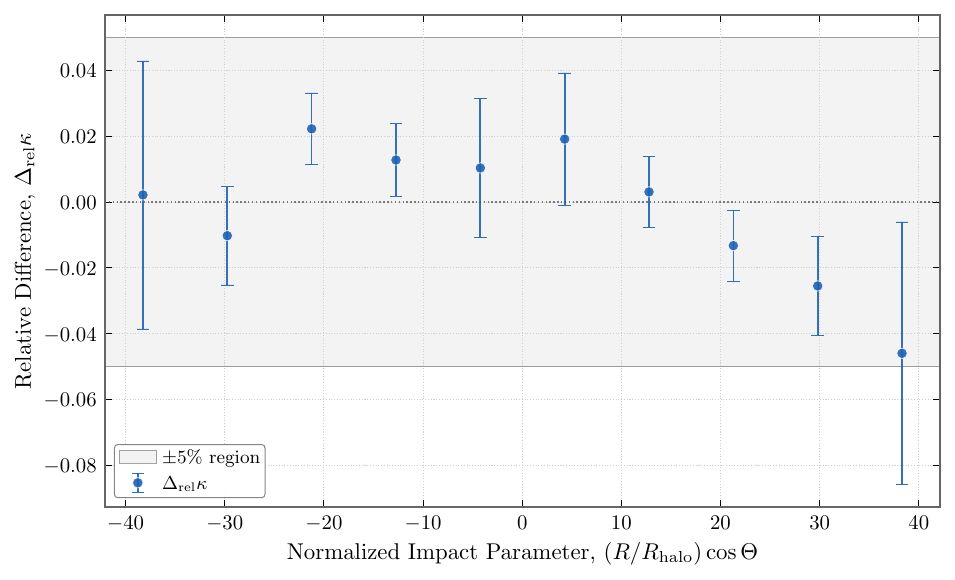}
        \caption{$\delrelk$, $10^{12.5}$--$10^{13.0}\hMsun$ halos.}
        \label{fig:delta_rel_k_halo_12.5-13.0}
    \end{subfigure}
    \caption{Doppler lensing convergence profiles around massive halos for three representative mass ranges: (a,b) $10^{11.5}$--$10^{12.0}\hMsun$, (c,d) $10^{12.0}$--$10^{12.5}\hMsun$, and (e,f) $10^{12.5}$--$10^{13.0}\hMsun$. Left panels show convergence $\kappa$ and difference $\Delta\kappa$; right panels show relative difference $\delrelk$ with $\pm5\%$ region highlighted. The excellent agreement across all these mass ranges contrasts with the significant discrepancies found in small voids.}
    \label{fig:k_halo}
\end{figure*}

Analysis of massive halos reveals consistent agreement between \gev and \scr across all mass ranges (Fig.~\ref{fig:k_halo}, Table~\ref{tab:halo_summary}). Mean absolute relative differences range from 1.6\% to 3.6\%, with 70--100\% of bins within $\pm5\%$ agreement. Only one bin across all mass ranges shows statistical significance ($|z|>2.81$, $\Delta_{\mathrm{rel}}\kappa=3.1\%$), confirming \scr's reliability in moderate-to-high density environments. The statistically significant bin is at mass range of $10^{12}$--$10^{12.5}\hMsun$, which contains the highest halo number density in our simulations, shows excellent agreement with a mean $|\Delta_{\mathrm{rel}}\kappa| = 2.4\%$ and 90\% of bins within $\pm5\%$.

\begin{table*}[!htbp]
    \centering
    \begin{tabular}{ccccc}
        \toprule
        \textbf{Halo Mass Range} & \textbf{Mean} & \textbf{Within $|\Delta_{\mathrm{rel}}\kappa|\le 0.05$} & \textbf{Mean} \\
        $\log_{10}(M/\hMsun)$ & \textbf{$|\Delta_{\mathrm{rel}}\kappa|$} & \textbf{ (\%)} & \textbf{$|z|$} \\
        \midrule
        $[11.5, 12)$ & 0.030 & 80.0 & 1.16 \\
        $[12, 12.5)$ & 0.024 & 90.0 & 1.33 \\
        $[12.5, 13)$ & 0.016 & 100.0 & 0.97 \\
        $[13, 13.5)$ & 0.027 & 90.0 & 1.27 \\
        $[13.5, 14)$ & 0.036 & 70.0 & 1.07 \\
        \bottomrule
    \end{tabular}        
    \caption{Summary of Doppler convergence analysis across different halo mass ranges from 21 realizations. The mean absolute relative difference ($|\Delta_{\mathrm{rel}}\kappa|$) shows the average discrepancy between Gevolution and Screening models, ranging from 1.6\% to 3.6\%. The "Within 5\%" column indicates the percentage of bins where the relative difference is $\le 5\%$, showing good agreement (70-100\% of bins). Mean $|z|$ values (0.97-1.33) indicate statistical significance levels.}
    \label{tab:halo_summary}
\end{table*}

\section{Conclusion}
\label{sec:conclusions}

The comparison of Doppler lensing convergence using \scr and \gev approaches reveals a dynamic scale-dependent performance across void sizes, highlighting the \scr method's computational efficiency while rigorously evaluating both methods' applicability. For small voids spanning 15 to 25\hMpc, the \scr approach underestimates approaching flows by 9.6\% and overestimates receding flows by 15.9\%, yielding an average absolute relative difference of 38.5\%. These biases, stemming from the method's linearized dynamics in high-density-contrast environments, are statistically significant. Medium voids of 25 to 35\hMpc show relatively much better agreement, with an average absolute relative difference of 9.5\% and minimal bias of about 1.9\%, demonstrating the \scr method's effectiveness for moderate-density environments ideal for statistical analyses. Large voids from 35 to 45\hMpc exhibit moderate errors, with an average absolute relative difference of 16.9\%, driven by challenges with density gradients and central instabilities. The persistent asymmetry across all ranges underscores the \scr method's radial velocity modeling challenges, but its efficiency in computation supports its strategic use where precision is less critical. Our halo analysis further confirms \scr's reliability in overdense environments, with mean differences of only 1.6--3.6\%.

In conclusion, our findings guide the selection of simulation frameworks based on scientific requirements. For precision studies where methodological accuracy is critical—such as testing $\Lambda$CDM deviations or constraining modified gravity—the \gev approach is essential due to its comprehensive treatment of nonlinear effects. For statistical analyses of halos and voids in surveys like Euclid or DESI, where computational efficiency enables larger sample sizes, \scr offers a practical alternative with methodological differences that may be acceptable for some applications.
Future efforts should validate these results with survey observations and develop refined simulation techniques for Doppler lensing measurements.

\section*{Acknowledgments}
We thank the anonymous referee for providing useful remarks that contributed to the final form of this paper. This work was supported by the Bangladesh University Grants Commission. We acknowledge the Bangladesh Research and Education Network (BdREN) for providing high-performance computing resources that have contributed to the research results reported within this paper.

\bibliographystyle{elsarticle-harv}
\bibliography{references}

@article{adamek2013,
  title = {General {{Relativistic N-body}} Simulations in the Weak Field Limit},
  author = {Adamek, Julian and Daverio, David and Durrer, Ruth and Kunz, Martin},
  year = {2013},
  month = nov,
  journal = {Physical Review D},
  volume = {88},
  number = {10},
  eprint = {1308.6524},
  primaryclass = {astro-ph},
  pages = {103527},
  issn = {1550-7998, 1550-2368},
  doi = {10.1103/PhysRevD.88.103527},
  urldate = {2025-01-09},
  archiveprefix = {arXiv},
  langid = {english}
}

@article{adamek2014,
  title = {N-Body Methods for Relativistic Cosmology},
  author = {Adamek, Julian and Durrer, Ruth and Kunz, Martin},
  year = {2014},
  month = dec,
  journal = {Classical and Quantum Gravity},
  volume = {31},
  number = {23},
  eprint = {1408.3352},
  primaryclass = {astro-ph, physics:gr-qc},
  pages = {234006},
  issn = {0264-9381, 1361-6382},
  doi = {10.1088/0264-9381/31/23/234006},
  urldate = {2024-08-13},
  archiveprefix = {arXiv}
}

@article{adamek2016,
  title = {General Relativity and Cosmic Structure Formation},
  author = {Adamek, Julian and Daverio, David and Durrer, Ruth and Kunz, Martin},
  year = {2016},
  month = apr,
  journal = {Nature Physics},
  volume = {12},
  number = {4},
  pages = {346--349},
  publisher = {Nature Publishing Group},
  issn = {1745-2481},
  doi = {10.1038/nphys3673},
  urldate = {2024-11-26},
  copyright = {2016 Springer Nature Limited},
  langid = {english}
}

@article{adamek2016a,
  title = {Gevolution: A Cosmological {{N-body}} Code Based on {{General Relativity}}},
  shorttitle = {Gevolution},
  author = {Adamek, Julian and Daverio, David and Durrer, Ruth and Kunz, Martin},
  year = {2016},
  month = jul,
  journal = {Journal of Cosmology and Astroparticle Physics},
  volume = {2016},
  number = {07},
  pages = {053},
  issn = {1475-7516},
  doi = {10.1088/1475-7516/2016/07/053},
  urldate = {2024-11-26},
  langid = {english}
}

@article{aghanim2020,
  title = {Planck 2018 Results. {{VI}}. {{Cosmological}} Parameters},
  author = {Aghanim, N. and Akrami, Y. and Ashdown, M. and Aumont, J. and Baccigalupi, C. and Ballardini, M. and Banday, A. J. and Barreiro, R. B. and Bartolo, N. and Basak, S. and Battye, R. and Benabed, K. and Bernard, J.-P. and Bersanelli, M. and Bielewicz, P. and Bock, J. J. and Bond, J. R. and Borrill, J. and Bouchet, F. R. and Boulanger, F. and Bucher, M. and Burigana, C. and Butler, R. C. and Calabrese, E. and Cardoso, J.-F. and Carron, J. and Challinor, A. and Chiang, H. C. and Chluba, J. and Colombo, L. P. L. and Combet, C. and Contreras, D. and Crill, B. P. and Cuttaia, F. and de Bernardis, P. and de Zotti, G. and Delabrouille, J. and Delouis, J.-M. and Valentino, E. Di and Diego, J. M. and Dor{\'e}, O. and Douspis, M. and Ducout, A. and Dupac, X. and Dusini, S. and Efstathiou, G. and Elsner, F. and En{\ss}lin, T. A. and Eriksen, H. K. and Fantaye, Y. and Farhang, M. and Fergusson, J. and {Fernandez-Cobos}, R. and Finelli, F. and Forastieri, F. and Frailis, M. and Fraisse, A. A. and Franceschi, E. and Frolov, A. and Galeotta, S. and Galli, S. and Ganga, K. and {G{\'e}nova-Santos}, R. T. and Gerbino, M. and Ghosh, T. and {Gonz{\'a}lez-Nuevo}, J. and G{\'o}rski, K. M. and Gratton, S. and Gruppuso, A. and Gudmundsson, J. E. and Hamann, J. and Handley, W. and Hansen, F. K. and Herranz, D. and Hildebrandt, S. R. and Hivon, E. and Huang, Z. and Jaffe, A. H. and Jones, W. C. and Karakci, A. and Keih{\"a}nen, E. and Keskitalo, R. and Kiiveri, K. and Kim, J. and Kisner, T. S. and Knox, L. and Krachmalnicoff, N. and Kunz, M. and {Kurki-Suonio}, H. and Lagache, G. and Lamarre, J.-M. and Lasenby, A. and Lattanzi, M. and Lawrence, C. R. and Jeune, M. Le and Lemos, P. and Lesgourgues, J. and Levrier, F. and Lewis, A. and Liguori, M. and Lilje, P. B. and Lilley, M. and Lindholm, V. and {L{\'o}pez-Caniego}, M. and Lubin, P. M. and Ma, Y.-Z. and {Mac{\'i}as-P{\'e}rez}, J. F. and Maggio, G. and Maino, D. and Mandolesi, N. and Mangilli, A. and {Marcos-Caballero}, A. and Maris, M. and Martin, P. G. and Martinelli, M. and {Mart{\'i}nez-Gonz{\'a}lez}, E. and Matarrese, S. and Mauri, N. and McEwen, J. D. and Meinhold, P. R. and Melchiorri, A. and Mennella, A. and Migliaccio, M. and Millea, M. and Mitra, S. and {Miville-Desch{\^e}nes}, M.-A. and Molinari, D. and Montier, L. and Morgante, G. and Moss, A. and Natoli, P. and {N{\o}rgaard-Nielsen}, H. U. and Pagano, L. and Paoletti, D. and Partridge, B. and Patanchon, G. and Peiris, H. V. and Perrotta, F. and Pettorino, V. and Piacentini, F. and Polastri, L. and Polenta, G. and Puget, J.-L. and Rachen, J. P. and Reinecke, M. and Remazeilles, M. and Renzi, A. and Rocha, G. and Rosset, C. and Roudier, G. and {Rubi{\~n}o-Mart{\'i}n}, J. A. and {Ruiz-Granados}, B. and Salvati, L. and Sandri, M. and Savelainen, M. and Scott, D. and Shellard, E. P. S. and Sirignano, C. and Sirri, G. and Spencer, L. D. and Sunyaev, R. and {Suur-Uski}, A.-S. and Tauber, J. A. and Tavagnacco, D. and Tenti, M. and Toffolatti, L. and Tomasi, M. and Trombetti, T. and Valenziano, L. and Valiviita, J. and Tent, B. Van and Vibert, L. and Vielva, P. and Villa, F. and Vittorio, N. and Wandelt, B. D. and Wehus, I. K. and White, M. and White, S. D. M. and Zacchei, A. and Zonca, A.},
  year = {2020},
  month = sep,
  journal = {Astronomy \& Astrophysics},
  volume = {641},
  eprint = {1807.06209},
  primaryclass = {astro-ph},
  pages = {A6},
  issn = {0004-6361, 1432-0746},
  doi = {10.1051/0004-6361/201833910},
  urldate = {2025-03-18},
  archiveprefix = {arXiv}
}

@article{amendola2018,
  title = {Cosmology and {{Fundamental Physics}} with the {{Euclid Satellite}}},
  author = {Amendola, Luca and Appleby, Stephen and Avgoustidis, Anastasios and Bacon, David and Baker, Tessa and Baldi, Marco and Bartolo, Nicola and Blanchard, Alain and Bonvin, Camille and Borgani, Stefano and Branchini, Enzo and Burrage, Clare and Camera, Stefano and Carbone, Carmelita and Casarini, Luciano and Cropper, Mark and de Rham, Claudia and Dietrich, Joerg P. and Porto, Cinzia Di and Durrer, Ruth and Ealet, Anne and Ferreira, Pedro G. and Finelli, Fabio and {Garcia-Bellido}, Juan and Giannantonio, Tommaso and Guzzo, Luigi and Heavens, Alan and Heisenberg, Lavinia and Heymans, Catherine and Hoekstra, Henk and Hollenstein, Lukas and Holmes, Rory and Horst, Ole and Hwang, Zhiqi and Jahnke, Knud and Kitching, Thomas D. and Koivisto, Tomi and Kunz, Martin and Vacca, Giuseppe La and Linder, Eric and March, Marisa and Marra, Valerio and Martins, Carlos and Majerotto, Elisabetta and Markovic, Dida and Marsh, David and Marulli, Federico and Massey, Richard and Mellier, Yannick and Montanari, Francesco and Mota, David F. and Nunes, Nelson J. and Percival, Will and Pettorino, Valeria and Porciani, Cristiano and Quercellini, Claudia and Read, Justin and Rinaldi, Massimiliano and Sapone, Domenico and Sawicki, Ignacy and Scaramella, Roberto and Skordis, Constantinos and Simpson, Fergus and Taylor, Andy and Thomas, Shaun and Trotta, Roberto and Verde, Licia and Vernizzi, Filippo and Vollmer, Adrian and Wang, Yun and Weller, Jochen and Zlosnik, Tom},
  year = {2018},
  month = dec,
  journal = {Living Reviews in Relativity},
  volume = {21},
  number = {1},
  eprint = {1606.00180},
  primaryclass = {astro-ph},
  pages = {2},
  issn = {2367-3613, 1433-8351},
  doi = {10.1007/s41114-017-0010-3},
  urldate = {2025-03-18},
  archiveprefix = {arXiv}
}

@misc{blas2011,
  title = {The {{Cosmic Linear Anisotropy Solving System}} ({{CLASS}}) {{II}}: {{Approximation}} Schemes},
  shorttitle = {The {{Cosmic Linear Anisotropy Solving System}} ({{CLASS}}) {{II}}},
  author = {Blas, Diego and Lesgourgues, Julien and Tram, Thomas},
  year = {2011},
  month = aug,
  eprint = {1104.2933},
  primaryclass = {astro-ph},
  doi = {10.1088/1475-7516/2011/07/034},
  urldate = {2025-03-18},
  archiveprefix = {arXiv}
}

@article{bonvin2008,
  title = {Effect of Peculiar Motion in Weak Lensing},
  author = {Bonvin, Camille},
  year = {2008},
  month = dec,
  journal = {Physical Review D},
  volume = {78},
  number = {12},
  pages = {123530},
  publisher = {American Physical Society},
  doi = {10.1103/PhysRevD.78.123530},
  urldate = {2025-01-21}
}

@article{bonvin2017,
  title = {Dipolar Modulation in the Size of Galaxies: The Effect of {{Doppler}} Magnification},
  shorttitle = {Dipolar Modulation in the Size of Galaxies},
  author = {Bonvin, Camille and Andrianomena, Sambatra and Bacon, David and Clarkson, Chris and Maartens, Roy and Moloi, Teboho and Bull, Philip},
  year = {2017},
  month = dec,
  journal = {Monthly Notices of the Royal Astronomical Society},
  volume = {472},
  number = {4},
  pages = {3936--3951},
  issn = {0035-8711},
  doi = {10.1093/mnras/stx2049},
  urldate = {2024-07-29}
}

@article{brilenkov2017,
  title = {Second-Order {{Cosmological Perturbations Engendered}} by {{Point-like Masses}}},
  author = {Brilenkov, Ruslan and Eingorn, Maxim},
  year = {2017},
  month = aug,
  journal = {The Astrophysical Journal},
  volume = {845},
  number = {2},
  eprint = {1703.10282},
  primaryclass = {astro-ph, physics:gr-qc, physics:hep-ph, physics:hep-th},
  pages = {153},
  issn = {0004-637X, 1538-4357},
  doi = {10.3847/1538-4357/aa81cd},
  urldate = {2024-08-13},
  archiveprefix = {arXiv}
}

@article{canay2021,
  title = {Scalar and Vector Perturbations in a Universe with Nonlinear Perfect Fluid},
  author = {Canay, Ezgi and Brilenkov, Ruslan and Eingorn, Maxim and Arapo{\u g}lu, A. Sava{\c s} and Zhuk, Alexander},
  year = {2021},
  month = mar,
  journal = {The European Physical Journal C},
  volume = {81},
  number = {3},
  eprint = {2011.05914},
  primaryclass = {gr-qc},
  pages = {246},
  issn = {1434-6044, 1434-6052},
  doi = {10.1140/epjc/s10052-021-09032-9},
  urldate = {2025-03-18},
  archiveprefix = {arXiv}
}

@article{davis1985,
  title = {The Evolution of Large-Scale Structure in a Universe Dominated by Cold Dark Matter},
  author = {Davis, M. and Efstathiou, G. and Frenk, C. S. and White, S. D. M.},
  year = {1985},
  month = may,
  journal = {The Astrophysical Journal},
  volume = {292},
  pages = {371--394},
  publisher = {IOP},
  issn = {0004-637X},
  doi = {10.1086/163168},
  urldate = {2025-03-18}
}

@article{eingorn2016,
  title = {First-Order {{Cosmological Perturbations Engendered}} by {{Point-like Masses}}},
  author = {Eingorn, Maxim},
  year = {2016},
  month = jul,
  journal = {The Astrophysical Journal},
  volume = {825},
  number = {2},
  eprint = {1509.03835},
  primaryclass = {astro-ph, physics:gr-qc, physics:hep-ph, physics:hep-th},
  pages = {84},
  issn = {0004-637X, 1538-4357},
  doi = {10.3847/0004-637X/825/2/84},
  urldate = {2024-08-13},
  archiveprefix = {arXiv}
}

@article{eingorn2016a,
  title = {Scalar and Vector Perturbations in a Universe with Discrete and Continuous Matter Sources},
  author = {Eingorn, Maxim and Kiefer, Claus and Zhuk, Alexander},
  year = {2016},
  month = sep,
  journal = {Journal of Cosmology and Astroparticle Physics},
  volume = {2016},
  number = {09},
  eprint = {1607.03394},
  primaryclass = {gr-qc},
  pages = {032--032},
  issn = {1475-7516},
  doi = {10.1088/1475-7516/2016/09/032},
  urldate = {2025-03-18},
  archiveprefix = {arXiv}
}

@article{eingorn2017,
  title = {Perfect Fluids with $\omega=\mathrm{const}$ as Sources of Scalar Cosmological Perturbations},
  author = {Eingorn, Maxim and Brilenkov, Ruslan},
  year = {2017},
  month = sep,
  journal = {Physics of the Dark Universe},
  volume = {17},
  eprint = {1509.08181},
  primaryclass = {gr-qc},
  pages = {63--67},
  issn = {22126864},
  doi = {10.1016/j.dark.2017.08.003},
  urldate = {2025-03-18},
  archiveprefix = {arXiv}
}

@article{eingorn2019,
  title = {Analytic Expressions for the Second-Order Scalar Perturbations in the \${\textbackslash}{{Lambda}}\${{CDM Universe}} within the Cosmic Screening Approach},
  author = {Eingorn, Maxim and Guran, N. Duygu and Zhuk, Alexander},
  year = {2019},
  month = dec,
  journal = {Physics of the Dark Universe},
  volume = {26},
  eprint = {1903.09024},
  primaryclass = {astro-ph, physics:gr-qc, physics:hep-ph, physics:hep-th},
  pages = {100329},
  issn = {22126864},
  doi = {10.1016/j.dark.2019.100329},
  urldate = {2024-08-13},
  archiveprefix = {arXiv}
}

@article{eingorn2022,
  title = {Screening vs. Gevolution: {{In}} Chase of a Perfect Cosmological Simulation Code},
  shorttitle = {Screening vs. Gevolution},
  author = {Eingorn, Maxim and Y{\"u}kselci, A. Emrah and Zhuk, Alexander},
  year = {2022},
  month = mar,
  journal = {Physics Letters B},
  volume = {826},
  pages = {136911},
  issn = {0370-2693},
  doi = {10.1016/j.physletb.2022.136911},
  urldate = {2024-07-29}
}

@article{eingorn2024,
  title = {Suppression of Matter Density Growth at Scales Exceeding the Cosmic Screening Length},
  author = {Eingorn, Maxim and Yilmaz, Ezgi and Y{\"u}kselci, A. Emrah and Zhuk, Alexander},
  year = {2024},
  month = may,
  journal = {Journal of Cosmology and Astroparticle Physics},
  volume = {2024},
  number = {05},
  eprint = {2307.06920},
  primaryclass = {gr-qc},
  pages = {083},
  issn = {1475-7516},
  doi = {10.1088/1475-7516/2024/05/083},
  urldate = {2025-04-09},
  archiveprefix = {arXiv}
}

@article{ema2022,
  title = {The Density Distributions of Cosmic Structures: Impact of the Local Environment on Weak-Lensing Convergence},
  shorttitle = {The Density Distributions of Cosmic Structures},
  author = {Ema, Sonia Akter and Hossen, Md Rasel and Bolejko, Krzysztof and Lewis, Geraint F},
  year = {2022},
  month = jan,
  journal = {Monthly Notices of the Royal Astronomical Society},
  volume = {509},
  number = {2},
  pages = {3004--3014},
  issn = {0035-8711},
  doi = {10.1093/mnras/stab3134},
  urldate = {2025-03-18}
}

@book{gorbunov2011,
  title = {Introduction to the {{Theory}} of the {{Early Universe}}},
  author = {Gorbunov, Dmitry S and Rubakov, Valery A},
  year = {2011},
  month = feb,
  publisher = {World Scientific Publishing Co Pte Ltd},
  langid = {english}
}

@article{hossen2022,
  title = {Ringing the Universe with Cosmic Emptiness: Void Properties through a Combined Analysis of Stacked Weak Gravitational and {{Doppler}} Lensing},
  shorttitle = {Ringing the Universe with Cosmic Emptiness},
  author = {Hossen, Md Rasel and Ema, Sonia Akter and Bolejko, Krzysztof and Lewis, Geraint F},
  year = {2022},
  month = jul,
  journal = {Monthly Notices of the Royal Astronomical Society},
  volume = {513},
  number = {4},
  pages = {5575--5587},
  issn = {0035-8711},
  doi = {10.1093/mnras/stac1247},
  urldate = {2024-07-29}
}

@article{lepori2020,
  title = {Weak-Lensing Observables in Relativistic {{N-body}} Simulations},
  author = {Lepori, Francesca and Adamek, Julian and Durrer, Ruth and Clarkson, Chris and Coates, Louis},
  year = {2020},
  month = sep,
  journal = {Monthly Notices of the Royal Astronomical Society},
  volume = {497},
  number = {2},
  eprint = {2002.04024},
  primaryclass = {astro-ph},
  pages = {2078--2095},
  issn = {0035-8711, 1365-2966},
  doi = {10.1093/mnras/staa2024},
  urldate = {2025-01-09},
  archiveprefix = {arXiv},
  langid = {english}
}

@book{mukhanov2005,
  title = {Physical {{Foundations}} of {{Cosmology}}},
  author = {Mukhanov, Viatcheslav},
  year = {2005},
  publisher = {Cambridge University Press}
}

@article{nadathur2019,
  title = {Beyond {{BAO}}: {{Improving}} Cosmological Constraints from {{BOSS}} Data with Measurement of the Void-Galaxy Cross-Correlation},
  shorttitle = {Beyond {{BAO}}},
  author = {Nadathur, Seshadri and Carter, Paul M. and Percival, Will J. and Winther, Hans A. and Bautista, Julian E.},
  year = {2019},
  month = jul,
  journal = {Physical Review D},
  volume = {100},
  pages = {023504},
  publisher = {APS},
  issn = {1550-79980556-2821},
  doi = {10.1103/PhysRevD.100.023504},
  urldate = {2025-03-19}
}

@article{nadathur2019b,
  title = {{{REVOLVER}}: {{REal-space VOid Locations}} from {{suVEy Reconstruction}}},
  shorttitle = {{{REVOLVER}}},
  author = {Nadathur, Seshadri and Carter, Paul M. and Percival, Will J. and Winther, Hans A. and Bautista, Julian E.},
  year = {2019},
  month = jul,
  journal = {Astrophysics Source Code Library},
  pages = {ascl:1907.023},
  urldate = {2025-06-27}
}

@book{peebles1980,
  title = {The Large-Scale Structure of the Universe},
  author = {Peebles, P. J. E.},
  year = {1980},
  month = jan,
  publisher = {Princeton University Press},
}

@misc{robertson2017,
  title = {Large {{Synoptic Survey Telescope Galaxies Science Roadmap}}},
  author = {Robertson, Brant E. and Banerji, Manda and Cooper, Michael C. and Davies, Roger and Driver, Simon P. and Ferguson, Annette M. N. and Ferguson, Henry C. and Gawiser, Eric and Kaviraj, Sugata and Knapen, Johan H. and Lintott, Chris and Lotz, Jennifer and Newman, Jeffrey A. and Norman, Dara J. and Padilla, Nelson and Schmidt, Samuel J. and Smith, Graham P. and Tyson, J. Anthony and Verma, Aprajita and Zehavi, Idit and Armus, Lee and Avestruz, Camille and Barrientos, L. Felipe and Bowler, Rebecca A. A. and Bremer, Malcom N. and Conselice, Christopher J. and Davies, Jonathan and Demarco, Ricardo and Dickinson, Mark E. and Galaz, Gaspar and Grazian, Andrea and Holwerda, Benne W. and Jarvis, Matt J. and Kasliwal, Vishal and Lacerna, Ivan and Loveday, Jon and Marshall, Phil and Merlin, Emiliano and Napolitano, Nicola R. and Puzia, Thomas H. and Robotham, Aaron and Salim, Samir and Sereno, Mauro and Snyder, Gregory F. and Stott, John P. and Tissera, Patricia B. and Werner, Norbert and Yoachim, Peter and Borne, Kirk D. and Collaboration, Members of the LSST Galaxies Science},
  year = {2017},
  month = aug,
  number = {arXiv:1708.01617},
  eprint = {1708.01617},
  primaryclass = {astro-ph},
  publisher = {arXiv},
  doi = {10.48550/arXiv.1708.01617},
  urldate = {2025-03-18},
  archiveprefix = {arXiv}
}

@article{springel2005,
  title = {Simulating the Joint Evolution of Quasars, Galaxies and Their Large-Scale Distribution},
  author = {Springel, Volker and White, Simon D. M. and Jenkins, Adrian and Frenk, Carlos S. and Yoshida, Naoki and Gao, Liang and Navarro, Julio and Thacker, Robert and Croton, Darren and Helly, John and Peacock, John A. and Cole, Shaun and Thomas, Peter and Couchman, Hugh and Evrard, August and Colberg, Joerg and Pearce, Frazer},
  year = {2005},
  month = jun,
  journal = {Nature},
  volume = {435},
  number = {7042},
  eprint = {astro-ph/0504097},
  pages = {629--636},
  issn = {0028-0836, 1476-4687},
  doi = {10.1038/nature03597},
  urldate = {2025-03-18},
  archiveprefix = {arXiv}
}

@article{weltman2020,
  title = {Fundamental {{Physics}} with the {{Square Kilometre Array}}},
  author = {Weltman, A. and Bull, P. and Camera, S. and Kelley, K. and Padmanabhan, H. and Pritchard, J. and Raccanelli, A. and {Riemer-S{\o}rensen}, S. and Shao, L. and Andrianomena, S. and Athanassoula, E. and Bacon, D. and Barkana, R. and Bertone, G. and Bonvin, C. and Bosma, A. and Br{\"u}ggen, M. and Burigana, C. and B{\oe}hm, C. and Calore, F. and Cembranos, J. A. R. and Clarkson, C. and Connors, R. M. T. and de la {Cruz-Dombriz}, {\'A} and Dunsby, P. K. S. and Fonseca, J. and Fornengo, N. and Gaggero, D. and Harrison, I. and Larena, J. and Ma, Y.-Z. and Maartens, R. and {M{\'e}ndez-Isla}, M. and Mohanty, S. D. and Murray, S. G. and Parkinson, D. and Pourtsidou, A. and Quinn, P. J. and Regis, M. and Saha, P. and Sahl{\'e}n, M. and Sakellariadou, M. and Silk, J. and Trombetti, T. and Vazza, F. and Venumadhav, T. and Vidotto, F. and {Villaescusa-Navarro}, F. and Wang, Y. and Weniger, C. and Wolz, L. and Zhang, F. and Gaensler, B. M.},
  year = {2020},
  journal = {Publications of the Astronomical Society of Australia},
  volume = {37},
  eprint = {1810.02680},
  primaryclass = {astro-ph},
  pages = {e002},
  issn = {1323-3580, 1448-6083},
  doi = {10.1017/pasa.2019.42},
  urldate = {2025-03-18},
  archiveprefix = {arXiv}
}

@article{2013PhRvL.110b1302B,
       author = {{Bolejko}, Krzysztof and {Clarkson}, Chris and {Maartens}, Roy and {Bacon}, David and {Meures}, Nikolai and {Beynon}, Emma},
        title = "{Antilensing: The Bright Side of Voids}",
      journal = {Physical Review Letters},
         year = 2013,
        month = jan,
       volume = {110},
       number = {2},
          eid = {021302},
        pages = {021302},
          doi = {10.1103/PhysRevLett.110.021302},
archivePrefix = {arXiv},
       eprint = {1209.3142},
 primaryClass = {astro-ph.CO},
       adsurl = {https://ui.adsabs.harvard.edu/abs/2013PhRvL.110b1302B}
}

@article{2014MNRAS.443.1900B,
       author = {{Bacon}, David J. and {Andrianomena}, Sambatra and {Clarkson}, Chris and {Bolejko}, Krzysztof and {Maartens}, Roy},
        title = "{Cosmology with Doppler lensing}",
      journal = {Monthly Notices of the Royal Astronomical Society},
         year = 2014,
        month = sep,
       volume = {443},
       number = {3},
        pages = {1900-1915},
          doi = {10.1093/mnras/stu1270},
archivePrefix = {arXiv},
       eprint = {1401.3694},
 primaryClass = {astro-ph.CO},
       adsurl = {https://ui.adsabs.harvard.edu/abs/2014MNRAS.443.1900B}
}

@article{2017MNRAS.472.3936B,
       author = {{Bonvin}, Camille and {Andrianomena}, Sambatra and {Bacon}, David and {Clarkson}, Chris and {Maartens}, Roy and {Moloi}, Teboho and {Bull}, Philip},
        title = "{Dipolar modulation in the size of galaxies: the effect of Doppler magnification}",
      journal = {Monthly Notices of the Royal Astronomical Society},
         year = 2017,
        month = dec,
       volume = {472},
       number = {4},
        pages = {3936-3951},
          doi = {10.1093/mnras/stx2049},
archivePrefix = {arXiv},
       eprint = {1610.05946},
 primaryClass = {astro-ph.CO},
       adsurl = {https://ui.adsabs.harvard.edu/abs/2017MNRAS.472.3936B}
}

@article{2008MNRAS.386.2101N,
       author = {{Neyrinck}, Mark C.},
        title = "{ZOBOV: a parameter-free void-finding algorithm}",
      journal = {Monthly Notices of the Royal Astronomical Society},
         year = 2008,
        month = jun,
       volume = {386},
       number = {4},
        pages = {2101-2109},
          doi = {10.1111/j.1365-2966.2008.13180.x},
archivePrefix = {arXiv},
       eprint = {0712.3049},
 primaryClass = {astro-ph},
       adsurl = {https://ui.adsabs.harvard.edu/abs/2008MNRAS.386.2101N}
}

@article{2012ApJ...761...44S,
       author = {{Sutter}, P.~M. and {Lavaux}, Guilhem and {Wandelt}, Benjamin D. and {Weinberg}, David H.},
        title = "{A Public Void Catalog from the SDSS DR7 Galaxy Redshift Surveys Based on the Watershed Transform}",
      journal = {Astrophysical Journal},
         year = 2012,
        month = dec,
       volume = {761},
       number = {1},
          eid = {44},
        pages = {44},
          doi = {10.1088/0004-637X/761/1/44},
archivePrefix = {arXiv},
       eprint = {1207.2524},
 primaryClass = {astro-ph.CO},
       adsurl = {https://ui.adsabs.harvard.edu/abs/2012ApJ...761...44S}
}

@article{2013ApJ...762..109B,
       author = {{Behroozi}, Peter S. and {Wechsler}, Risa H. and {Wu}, Hao-Yi},
        title = "{The ROCKSTAR Phase-space Temporal Halo Finder and the Velocity Offsets of Cluster Cores}",
      journal = {Astrophysical Journal},
         year = 2013,
        month = jan,
       volume = {762},
       number = {2},
          eid = {109},
        pages = {109},
          doi = {10.1088/0004-637X/762/2/109},
archivePrefix = {arXiv},
       eprint = {1110.4372},
 primaryClass = {astro-ph.CO},
       adsurl = {https://ui.adsabs.harvard.edu/abs/2013ApJ...762..109B}
}

@article{2015JCAP...08..028B,
       author = {{Barreira}, Alexandre and {Cautun}, Marius and {Li}, Baojiu and {Baugh}, Carlton M. and {Pascoli}, Silvia},
        title = "{Weak lensing by voids in modified lensing potentials}",
      journal = {Journal of Cosmology and Astroparticle Physics},
         year = 2015,
        month = aug,
       volume = {2015},
       number = {8},
          eid = {028},
        pages = {028},
          doi = {10.1088/1475-7516/2015/08/028},
archivePrefix = {arXiv},
       eprint = {1505.05809},
 primaryClass = {astro-ph.CO},
       adsurl = {https://ui.adsabs.harvard.edu/abs/2015JCAP...08..028B}
}

@article{2015PhRvD..92h3531P,
       author = {{Pisani}, Alice and {Sutter}, P.~M. and {Hamaus}, Nico and {Alizadeh}, Esfandiar and {Biswas}, Rahul and {Wandelt}, Benjamin D. and {Hirata}, Christopher M.},
        title = "{Counting voids to probe dark energy}",
      journal = {Physical Review D},
         year = 2015,
        month = oct,
       volume = {92},
       number = {8},
          eid = {083531},
        pages = {083531},
          doi = {10.1103/PhysRevD.92.083531},
archivePrefix = {arXiv},
       eprint = {1503.07690},
 primaryClass = {astro-ph.CO},
       adsurl = {https://ui.adsabs.harvard.edu/abs/2015PhRvD..92h3531P}
}

@article{2016ApJ...820L...7S,
       author = {{Sahl{\'e}n}, Martin and {Zubeld{\'\i}a}, {\'I}{\~n}igo and {Silk}, Joseph},
        title = "{Cluster-Void Degeneracy Breaking: Dark Energy, Planck, and the Largest Cluster and Void}",
      journal = {Astrophysics Letters},
         year = 2016,
        month = mar,
       volume = {820},
       number = {1},
          eid = {L7},
        pages = {L7},
          doi = {10.3847/2041-8205/820/1/L7},
archivePrefix = {arXiv},
       eprint = {1511.04075},
 primaryClass = {astro-ph.CO},
       adsurl = {https://ui.adsabs.harvard.edu/abs/2016ApJ...820L...7S}
}

@article{2019BAAS...51c..40P,
       author = {{Pisani}, Alice and {Massara}, Elena and {Spergel}, David N. and {Alonso}, David and {Baker}, Tessa and {Cai}, Yan-Chuan and {Cautun}, Marius and {Davies}, Christopher and {Demchenko}, Vasiliy and {Dor{\'e}}, Olivier and {Goulding}, Andy and {Habouzit}, M{\'e}lanie and {Hamaus}, Nico and {Hawken}, Adam and {Hirata}, Christopher M. and {Ho}, Shirley and {Jain}, Bhuvnesh and {Kreisch}, Christina D. and {Marulli}, Federico and {Padilla}, Nelson and {Pollina}, Giorgia and {Sahl{\'e}n}, Martin and {Sheth}, Ravi K. and {Somerville}, Rachel and {Szapudi}, Istvan and {van de Weygaert}, Rien and {Villaescusa-Navarro}, Francisco and {Wandelt}, Benjamin D. and {Wang}, Yun},
        title = "{Cosmic voids: a novel probe to shed light on our Universe}",
      journal = {Bulletin of the American Astronomical Society},
         year = 2019,
        month = may,
       volume = {51},
       number = {3},
          eid = {40},
        pages = {40},
archivePrefix = {arXiv},
       eprint = {1903.05161},
 primaryClass = {astro-ph.CO},
       adsurl = {https://ui.adsabs.harvard.edu/abs/2019BAAS...51c..40P}
}

@article{2022MNRAS.509.5142H,
       author = {{Hossen}, Md Rasel and {Ema}, Sonia Akter and {Bolejko}, Krzysztof and {Lewis}, Geraint F.},
        title = "{Mapping the cosmic mass distribution with stacked weak gravitational lensing and Doppler lensing}",
      journal = {Monthly Notices of the Royal Astronomical Society},
         year = 2022,
        month = feb,
       volume = {509},
       number = {4},
        pages = {5142-5154},
          doi = {10.1093/mnras/stab3292},
archivePrefix = {arXiv},
       eprint = {2111.05439},
 primaryClass = {astro-ph.CO},
       adsurl = {https://ui.adsabs.harvard.edu/abs/2022MNRAS.509.5142H}
}

@article{2009ApJ...696L..10L,
       author = {{Lee}, Jounghun and {Park}, Daeseong},
        title = "{Constraining the Dark Energy Equation of State with Cosmic Voids}",
      journal = {Astrophysics Letters},
         year = 2009,
        month = may,
       volume = {696},
       number = {1},
        pages = {L10-L12},
          doi = {10.1088/0004-637X/696/1/L10},
archivePrefix = {arXiv},
       eprint = {0704.0881},
 primaryClass = {astro-ph},
       adsurl = {https://ui.adsabs.harvard.edu/abs/2009ApJ...696L..10L}
}

@article{2019MNRAS.490.3573F,
       author = {{Fang}, Y. and {Hamaus}, N. and {Jain}, B. and {Pandey}, S. and {Pollina}, G. and {S{\'a}nchez}, C. and {Kov{\'a}cs}, A. and {Chang}, C. and {Carretero}, J. and {Castander}, F.~J. and {Choi}, A. and {Crocce}, M. and {DeRose}, J. and {Fosalba}, P. and {Gatti}, M. and {Gazta{\~n}aga}, E. and {Gruen}, D. and {Hartley}, W.~G. and {Hoyle}, B. and {MacCrann}, N. and {Prat}, J. and {Rau}, M.~M. and {Rykoff}, E.~S. and {Samuroff}, S. and {Sheldon}, E. and {Troxel}, M.~A. and {Vielzeuf}, P. and {Zuntz}, J. and {Annis}, J. and {Avila}, S. and {Bertin}, E. and {Brooks}, D. and {Burke}, D.~L. and {Carnero Rosell}, A. and {Carrasco Kind}, M. and {Cawthon}, R. and {da Costa}, L.~N. and {De Vicente}, J. and {Desai}, S. and {Diehl}, H.~T. and {Dietrich}, J.~P. and {Doel}, P. and {Everett}, S. and {Evrard}, A.~E. and {Flaugher}, B. and {Frieman}, J. and {Garc{\'\i}a-Bellido}, J. and {Gerdes}, D.~W. and {Gruendl}, R.~A. and {Gutierrez}, G. and {Hollowood}, D.~L. and {James}, D.~J. and {Jarvis}, M. and {Kuropatkin}, N. and {Lahav}, O. and {Maia}, M.~A.~G. and {Marshall}, J.~L. and {Melchior}, P. and {Menanteau}, F. and {Miquel}, R. and {Palmese}, A. and {Plazas}, A.~A. and {Romer}, A.~K. and {Roodman}, A. and {Sanchez}, E. and {Serrano}, S. and {Sevilla-Noarbe}, I. and {Smith}, M. and {Soares-Santos}, M. and {Sobreira}, F. and {Suchyta}, E. and {Swanson}, M.~E.~C. and {Tarle}, G. and {Thomas}, D. and {Vikram}, V. and {Walker}, A.~R. and {Weller}, J. and {DES Collaboration}},
        title = "{Dark Energy Survey year 1 results: the relationship between mass and light around cosmic voids}",
      journal = {Monthly Notices of the Royal Astronomical Society},
         year = 2019,
        month = dec,
       volume = {490},
       number = {3},
        pages = {3573-3587},
          doi = {10.1093/mnras/stz2805},
archivePrefix = {arXiv},
       eprint = {1909.01386},
 primaryClass = {astro-ph.CO},
       adsurl = {https://ui.adsabs.harvard.edu/abs/2019MNRAS.490.3573F}
}

@article{2012PhRvD..85f3512G,
       author = {{Green}, Stephen R. and {Wald}, Robert M.},
        title = "{Newtonian and relativistic cosmologies}",
      journal = {Physical Review D},
         year = 2012,
        month = mar,
       volume = {85},
       number = {6},
          eid = {063512},
        pages = {063512},
          doi = {10.1103/PhysRevD.85.063512},
archivePrefix = {arXiv},
       eprint = {1111.2997},
 primaryClass = {gr-qc},
       adsurl = {https://ui.adsabs.harvard.edu/abs/2012PhRvD..85f3512G}
}

@article{2022arXiv220107241M,
       author = {{Moresco}, Michele and {Amati}, Lorenzo and {Amendola}, Luca and {Birrer}, Simon and {Blakeslee}, John P. and {Cantiello}, Michele and {Cimatti}, Andrea and {Darling}, Jeremy and {Della Valle}, Massimo and {Fishbach}, Maya and {Grillo}, Claudio and {Hamaus}, Nico and {Holz}, Daniel and {Izzo}, Luca and {Jimenez}, Raul and {Lusso}, Elisabeta and {Meneghetti}, Massimo and {Piedipalumbo}, Ester and {Pisani}, Alice and {Pourtsidou}, Alkistis and {Pozzetti}, Lucia and {Quartin}, Miguel and {Risaliti}, Guido and {Rosati}, Piero and {Verde}, Licia},
        title = "{Unveiling the Universe with Emerging Cosmological Probes}",
      journal = {arXiv e-prints},
         year = 2022,
        month = jan,
          eid = {arXiv:2201.07241},
        pages = {arXiv:2201.07241},
archivePrefix = {arXiv},
       eprint = {2201.07241},
 primaryClass = {astro-ph.CO},
       adsurl = {https://ui.adsabs.harvard.edu/abs/2022arXiv220107241M}
}

@article{2020JCAP...12..023H,
       author = {{Hamaus}, Nico and {Pisani}, Alice and {Choi}, Jin-Ah and {Lavaux}, Guilhem and {Wandelt}, Benjamin D. and {Weller}, Jochen},
        title = "{Precision cosmology with voids in the final BOSS data}",
      journal = {Journal of Cosmology and Astroparticle Physics},
         year = 2020,
        month = dec,
       volume = {2020},
       number = {12},
          eid = {023},
        pages = {023},
          doi = {10.1088/1475-7516/2020/12/023},
archivePrefix = {arXiv},
       eprint = {2007.07895},
 primaryClass = {astro-ph.CO},
       adsurl = {https://ui.adsabs.harvard.edu/abs/2020JCAP...12..023H}
}

@article{2022MNRAS.510.2980A,
       author = {{Anbajagane}, Dhayaa and {Aung}, Han and {Evrard}, August E. and {Farahi}, Arya and {Nagai}, Daisuke and {Barnes}, David J. and {Cui}, Weiguang and {Dolag}, Klaus and {McCarthy}, Ian G. and {Rasia}, Elena and {Yepes}, Gustavo},
        title = "{Galaxy velocity bias in cosmological simulations: towards per cent-level calibration}",
      journal = {Monthly Notices of the Royal Astronomical Society},
         year = 2022,
        month = feb,
       volume = {510},
       number = {2},
        pages = {2980-2997},
          doi = {10.1093/mnras/stab3587},
archivePrefix = {arXiv},
       eprint = {2110.01683},
 primaryClass = {astro-ph.CO},
       adsurl = {https://ui.adsabs.harvard.edu/abs/2022MNRAS.510.2980A}
}

\end{document}